%% file: main.tex
\documentclass{elsarticle}

\usepackage{amsthm}
\usepackage{graphicx} 
\usepackage{amssymb}
\usepackage{amsmath}
\usepackage{times}
\usepackage{enumerate}
\usepackage{todonotes}
\usepackage{multirow}
\usepackage{hyperref}
\hypersetup{%
	colorlinks=true,           
	allcolors=blue!70!black,   
	pdfstartview=Fit,          
	breaklinks=true,
	pdfauthor={B. Bonakdarpour and S. Sheinvald},
	pdftitle={Finite-Word Hyperlanguages}}

\usepackage[vlined,boxruled,linesnumbered,noend]{algorithm2e}  
\SetKwInput{KwInput}{Input}                
\SetKwInput{KwOutput}{Output}

\begin{document}
\input{macros}

\title{Finite-Word Hyperlanguages}

\author[1]{Borzoo Bonakdarpour}\ead{borzoo@msu.edu}

\author[2]{Sarai Sheinvald}\ead{sarai@braude.ac.il}

\address[1]{Department of Computer Science and Engineering, Michigan State University, USA}

\address[2]{Department of Software Engineering, ORT Braude College, Israel}
\input{abs}
\date{}
\maketitle


\input{intro}
\input{preliminaries}
\input{HRE}

\input{NFH}
\input{nfh_properties}

\input{nonemptiness}

\input{decision_procedures}
\input{related}
\input{discussion}

\bibliographystyle{splncs04}
\bibliography{biblio2}



 


\end{document}

%% file: macros.tex
\newcommand{\f}{\varphi}
\newcommand{\g}{\psi}

\newcommand{\globally}{\textsf {G}\, }
\newcommand{\eventually}{\textsf{F}\, }
\newcommand{\weakuntil}{\textsf{W}\, }
\newcommand{\until}{\, \textsf{U}\, }
\newcommand{\releases}{\, \textsf{V}\, }
\newcommand{\nextt}{\textsf{X}\, }
\newcommand{\true}{\textbf{\textit{tt}}}
\newcommand{\false}{\textbf{\textit{ff}}}

\newcommand{\AP}{\mathsf{AP}}
\newcommand{\zug}[1]{\langle #1 \rangle}
\newcommand{\tuple}[1]{\langle #1 \rangle}
\newcommand{\A}{\mathcal{A}}
\newcommand{\B}{\mathcal{B}}
\newcommand{\lang}[1]{\mathcal{L}(#1)}
\newcommand{\hlang}[1]{\mathfrak{L}(#1)}
\newcommand{\hl}{\mathfrak{L}}
\newcommand{\K}{\mathcal{K}}
\newcommand{\M}{\mathcal{M}}
\newcommand{\D}{\mathcal{D}}
\newcommand{\U}{\mathcal{~U~}}
\newcommand{\lstar}{\textsc{L}^*}
\newcommand{\ceil}[1]{\lceil #1 \rceil}
\newcommand{\nfhef}{\textrm{NFH}_{\exists\forall}}
\newcommand{\nfhfe}{\textrm{NFH}_{\forall\exists}}
\newcommand{\nfhe}{\textrm{NFH}_{\exists}}
\newcommand{\nfhf}{\textrm{NFH}_{\forall}}
\newcommand{\hreef}{\textrm{HRE}_{\exists\forall}}
\newcommand{\hrefe}{\textrm{HRE}_{\forall\exists}}
\newcommand{\hree}{\textrm{HRE}_{\exists}}
\newcommand{\hrefall}{\textrm{HRE}_{\forall}}

\newcommand{\wass}[2]{{\bi #1}_{#2}}
\newcommand{\assw}[2]{{#1}_{\bi #2}}

\newcommand{\comp}[1]{\textsf{\small #1}}

\newcommand{\borzoo}[1]{\textcolor{red}{\bf #1}}
\newcommand{\alphabet}{\Sigma}

\newcommand{\naturals}{\mathbb{N}}
\newcommand{\N}{\naturals}

\newcommand{\bi}[1]{\textbf{\textit #1}} 

\newdefinition{example}{Example}
\newtheorem{theorem}{Theorem}
\newdefinition{definition}{Definition}

\newcommand{\quant}{\mathbb{Q}}
\newcommand{\stam}[1]{}
\newcommand{\zip}{\mathsf{zip}}
\newcommand{\unzip}{\mathsf{unzip}}
\newcommand{\row}{\mathsf{row}}
\newcommand{\range}{\mathsf{range}}

%% file: abs.tex
\begin{abstract}

{\em Formal languages} are in the core of models of computation and their behavior. A 
rich family of models for many classes of languages have been widely studied. 
{\em Hyperproperties} lift conventional trace-based languages from a set of execution 
traces to a set of sets of executions. Hyperproperties have been shown to be a 
powerful formalism for expressing and reasoning about information-flow security policies 
and important properties of cyber-physical systems.
Although there is an extensive body of work on formal-language representation of trace properties, we currently lack such a general characterization for hyperproperties.

We introduce {\em hyperlanguages} over finite words and models for expressing them. 
Essentially, these models express multiple words by using assignments to quantified {\em 
word variables}.
Relying on the standard models for regular languages, we propose {\em hyperregular expressions} and {\em finite-word hyperautomata (NFH)}, for modeling the class of {\em regular hyperlanguages}.
We demonstrate the ability of regular hyperlanguages to express 
hyperproperties for finite traces. We explore the closure properties and the complexity of 
the fundamental decision problems such as nonemptiness, universality, membership, and 
containment for various fragments of NFH. 

\stam{We then explore the fundamental properties of regular hyperlanguages. 
We show that NFH are closed under the Boolean operations, and that while nonemptiness of NFH is undecidable in general, it is decidable for several fragments of NFH. We further show the decidability of the membership problem for finite sets and regular languages for NFH, as well as for the universality and containment problem for several fragments of NFH.
}


\end{abstract}

%% file: intro.tex
\section{Introduction}

Formal languages, along with the models that express them, are in the core of modeling, 
specification, and verification of computing systems. Execution traces are formally 
described as words, and various families of automata are used for modeling systems of 
different types. {\em Regular languages} are a classic formalism for finite traces and when 
the traces are infinite, {\em $\omega$-regular languages} are used.

There are well-known connections between specification logics and formal languages. For example, 
LTL~\cite{p77} formulas can be translated to $\omega$-regular expressions, and CTL$^*$~\cite{eh86} 
formulas can be expressed using tree automata. 
Accordingly, many verification techniques that exploit these relations have been developed. For instance, in the automata-theoretic approach to verification~\cite{vw86,VW94}, the model-checking problem is reduced to checking the nonemptiness of the product automaton of the model and the complement of the specification.

{\em Hyperproperties}~\cite{cs10} generalize the traditional trace properties~\cite{as85} to {\em system properties}, i.e., a set of sets of traces. A hyperproperty prescribes how the system should behave in its entirety and not just based on its individual executions. 
Hyperproperties have been shown to be a powerful tool for expressing and reasoning about 
information-flow security policies~\cite{cs10} and important properties of cyber-physical 
systems~\cite{wzbp19} such as {\em sensitivity} and {\em robustness}, as well as consistency 
conditions in distributed computing such as
{\em linearizability}~\cite{bss18}. 
While different types of logics have been suggested for expressing hyperproperties, their formal-language counterparts and the  models that express them are currently missing. 

In this paper, we establish a formal-language theoretical framework for {\em 
hyperlanguages}, that are sets of sets of words, which we term {\em 
hyperwords}. Our framework is based on an underlying standard automata model for formal 
languages, augmented with quantified {\em word variables} that are assigned 
words from a set of words in the hyperlanguage. This formalism is in line with 
logics for hyperproperties (e.g., HyperLTL~\cite{cfkmrs14} and 
HyperPCTL~\cite{ab18,abbd20-atva}).
These logics express the behavior of infinite trace systems. However, a basic formal model for expressing general hyperproperties for finite 
words has not been defined yet. 

To begin with the basics, we focus this paper on a regular type of 
hyperlanguages of sets consisting of finite words, which we call {\em 
	regular hyperlanguages}. 
The models we introduce and study are based on the standard models for regular 
languages, namely regular expressions and finite-word automata.

\subsection{Motivation and Applications}

Hyperlanguages based on finite words have many practical applications. 
Let us first explain the idea of hyperlanguages with two examples.

\begin{example}\label{example:intro_1}
Consider the following {\em hyperregular expression} (HRE) over the alphabet $\{a\}$.
$$ r_1 = \forall x.\exists y.\underbrace{\Big(\{a_x, a_y\} ^*\{\#_x, a_y\}^* 
\Big)}_{\hat r_1}
$$
The HRE $r_1$ uses two word variables $x$ and $y$, which are assigned words 
from a hyperword. The  HRE $r_1$ contains an underlying regular expression 
$\hat r_1$, whose alphabet is $(\{a\}\cup\{\#\})^{\{x,y\}}$, and whose 
(regular) language describes different word assignments to $x$ and $y$, where 
$\#$ is used for padding at the end if the words assigned to $x$ and $y$ are of 
different lengths. In a word in the language of $\hat r_1$, 
the $i$'th letter describes both $i$'th letters in the words assigned to $x$ 
and $y$. For example, the word $\{a_x,a_y\}\{a_x,a_y\}\{\#_x, a_y\}$ 
describes the assignment $x\mapsto aa, y\mapsto aaa$.  
The regular expression ${\hat r_1}$ requires that the word assigned to $y$ be 
longer than the word assigned to $x$. 
The {\em quantification condition} $\forall x .\exists y$ of $r_1$ requires that for every word in a hyperword $S$ in the hyperlanguage of  
$r_1$, there exists a longer word in $S$. This holds iff $S$ contains 
infinitely many words. Therefore, the hyperlanguage of $r_1$ is the set of all 
infinite hyperwords over $\{a\}$. \qed
\end{example}

\begin{example}

Path planning objectives for robotic systems often stipulate the {\em existence} of one or more {\em 
finite} paths that stand out from {\em all} other paths.
For example, robotics applications are often concerned with finding the shortest path that reaches 
a goal $g$, starting from an initial location $i$. The shortest path 
requirement can be expressed by the following HRE over an alphabet $\alphabet$: 
$$r_2 = \exists x.\forall y. \{i_x, i_y\}\{\bar{g}_x, \bar{g}_y\}^*\Big(\{g_x, 
\bar{g}_y\}\mid \{g_x, {g}_y\}\Big)\{\#_x, \$_y\}^*$$
where $\bar{g} \in \alphabet - \{g\}$ and $\$ \in \alphabet$. That is, there exists a 
path $x$ that is shorter than any other path $y$ in reaching $g$.

Another interesting application in robotics is in {\em adversarial} settings, where some robots may 
interfere (e.g., act as moving obstacles) with a set of controllable robots. In this scenario, given any 
behavior of the adversarial robots, the controllable robots should be able to achieve their operation 
objectives.
This specification is in general of the following form:
$$
r_3 = \underbrace{\forall x_1.\forall x_2 \ldots \forall x_n}_{\text{advarsaries}}.\underbrace{\exists 
y_1.\exists 
y_2 \ldots \exists y_m}_{\text{controllable}}.\hat{r}
$$ 
where words $x_1 \cdots x_n$ express the behavior of the adversaries, words $y_1 \cdots y_m$ 
describe the behavior of the controllable robots and regular expression $\hat{r}$ specifies the control 
objectives.
\qed
\end{example}

\subsection{Contributions}

Although there is an ongoing line of research on model-checking 
hyperproperties~\cite{frs15,bf18,cfst19}, the work on finite-trace 
hyperproperties is limited to~\cite{fht19}, where the authors construct a 
finite-word representation for the class of regular $k$-safety hyperproperties. 
We make the following contributions:

\input{table}

%
%
\begin{itemize}
\item Introduce regular hyperlanguages and HREs, and demonstrate the ability of HREs to express 
important information-flow security policies such as different variations of {\em 
noninterference}~\cite{gm82} and 
{\em observational determinism}~\cite{zm03}.

\item Present {\em nondeterministic finite-word hyperautomata} (NFH), an automata-based model for expressing regular hyperlanguages.

\item Conduct a comprehensive study of the properties of regular hyperlanguages (see 
Table~\ref{tab:results}):

\begin{itemize}
	\item We show that regular hyperlanguages are {\em closed} under union, intersection, and 
complementation.

\item We consider the {\em nonemptiness} problem for NFH:

\begin{itemize} 
	
	\item We prove that the nonemptiness problem is in general 
undecidable for NFH. 

\item However, for the alternation-free fragments (which only 
allow one type of quantifier), as well as for the $\exists\forall$ fragment (in 
which the quantification condition is limited to a sequence of $\exists$ 
quantifiers followed by a sequence of $\forall$ 
quantifiers), nonemptiness is decidable. 

\item As another positive result in the area of nonemptiness, we show that the {\em bounded nonemptiness problem}, in which we decide whether an NFH accepts a hyperword of bounded size, is \comp{PSPACE-complete}.

\item We consider the construction of HRE and NFH with {\em wild card letters}, which allow expressing the assignment to only a subset of the variables, by assigning a wild card letter to the rest of the variables. We show that adding wild cards does not alter the complexity of the nonemptiness for the alternation-free fragments, while it does increase the complexity of this problem for the $\exists\forall$ fragment.

\item We describe a semi-algorithm for deciding the nonemptiness of NFH with a $\forall\exists$ quantification condition. The procedure begins with the largest potential hyperword, and iteratively prunes it in a consistent way in case it is not accepted. Since the problem is undecidable, there are inputs for which our semi-algorithm does not halt. However, in case it does halt, it is guaranteed to return a correct answer. Since $\forall\exists$ is a useful fragment, our procedure can be a useful tool. 

\end{itemize}

\item We study the {\em universality}, {\em membership} and {\em containment} problems. These results are 
aligned with the complexity of HyperLTL model checking for tree-shaped and 
general Kripke structures~\cite{bf18}. This shows that the complexity results in~\cite{bf18} mainly stem from the nature of quantification 
over finite words and depend on neither the full power of the temporal operators 
nor the infinite nature of HyperLTL semantics.

\end{itemize}
\end{itemize}

\stam{
\begin{itemize}
 \item As mentioned earlier, we first introduce the notion of HRE and 
demonstrate the ability of HREs to express important information-flow security 
policies such as different variations of {noninteference}~\cite{gm82} and 
{\em observational determinism}~\cite{zm03}.

\item We show that a regular hyperlanguage $\hl$ can also be expressed by a 
finite-word automaton (NFA) for the underlying regular expression of an HRE $r$ 
for $\hl$, augmented with the qauntification condition of $r$. We call this 
model a {\em finite-word hyperautomaton} (NFH).

\item Using NFH, we proceed to conduct a comprehensive study of properties of 
regular hyperlanguages (see Table~\ref{tab:results}). In particular, we show 
that regular hyperlanguages are {\em closed} under union, intersection, and 
complementation. 

\item We also prove that the {\em nonemptiness} problem is in general 
undecidable for NFH. However, for the alternation-free fragments (which only 
allow one type of quantifier), as well as for the $\exists\forall$ fragment (in 
which the quantification condition is limited to a sequence of $\exists$ 
quantifiers followed by a sequence of $\forall$ 
quantifiers), nonemptiness is decidable (see Table~\ref{tab:results}). We also 
study the {\em membership} and {\em inclusion} problems. These results are 
aligned with the complexity of HyperLTL model checking for tree-shaped and 
general Kripke structures~\cite{bf18}. This shows that, surprisingly, the 
complexity results in~\cite{bf18} mainly stem from the nature of quantification 
over finite words and depend on neither the full power of the temporal operators 
nor the infinite nature of HyperLTL semantics.

\item Finally, we lift the notion to hyperwords consisting of infinite words to 
{\em B{\"u}chi hyperautomata}, and show that using the standard construction for 
LTL as basis, the entire class of HyperLTL can be translated to B{\"u}chi 
hyperautomata. The exponential blow-up involved in the construction, along with 
the complexity results for the various decision problems are in line with the 
results on satisfiability and model-checking of HyperLTL~\cite{fh16} and its 
various fragments. Thus, B{\"u}chi hyperautomata are shown to be optimal, 
complexity-wise, for handling HyperLTL.

\end{itemize}
}

\paragraph{Comparison to the conference version} This article substantially extends the results of 
our original conference submission \cite{bs21} by the following new contributions. 
\begin{itemize}
\item An upper and lower bound of the bounded nonempitness problem.
\item Upper and lower bounds for the nonemptiness problem for the various fragments of NFH in the presence of wild-card letters.
\item A semi-algorithm for deciding the nonemptiness for the $\forall\exists$ fragment.
\item A detailed discussion on related work. 
\end{itemize}
\stam{
Our results on decidability and in particular 
\comp{PSPACE-completeness} of the $k$-nonemptiness problem (i.e., 
Theorem~\ref{thm:bounded.pspace.complete}) is new. We also introduce a semi-algorithm for 
deciding nonemptiness for the $\forall \exists$ fragment. The notion of NFH with wild cards as well 
as our results in Theorems~\ref{thm:nfhe.nfha.star.nonemptiness} 
and~\ref{thm:nfhef.star.nonemptiness} are 
also added to the original conference submission. We have also added a detailed discussion on related 
work. }
In summary, the material in Sections~\ref{sec:bounded},~\ref{sec:semialgo}~\ref{sec:wildcard}, 
and~\ref{sec:related} is all new. Finally, all proof sketches are now extended to revised and detailed full 
proofs.

\subsection{Organization} 

The rest of the paper is organized as follows. 
Preliminary concepts are presented in Section~\ref{sec:prelim}. We introduce 
the notion of HRE and NFH in Sections~\ref{sec:hre} and~\ref{sec:nfh}, while 
their properties and our complexity results are studied in Sections~\ref{sec:nfh_closure}, 
\ref{sec:nonemptiness}, and~\ref{sec:decproc}. Related work is discussed in Section~\ref{sec:related}. 
Finally, we make concluding remarks and discuss future work in Section~\ref{sec:concl}.

%% file: table.tex
\begin{table}[t]
\begin{center}
\begin{tabular}{|c||c|c|}

\multicolumn{1}{c}{\bf Property} & \multicolumn{2}{c}{\bf Result}\\
\hline\hline
Closure &  \multicolumn{2}{c|}{Complementation, Union, Intersection 
(Theorems~\ref{thm:nfh.complementation},~\ref{thm:nfh.union},~\ref{thm:nfh.intersection})} \\
\hline

\multirow{4}{*}{Nonemptiness}& $\forall\exists\exists$ & Undecidable 
(Theorem~\ref{thm:nfh.nonemptiness})\\ 
& $\exists^*~/~\forall^*~/~\exists^{*\star}~/~\forall^{*\star}$ & \comp{NL-complete} 
(Theorems~\ref{thm:nfhe.nfha.nonemptiness},~\ref{thm:nfhe.nfha.star.nonemptiness})\\ 
& $\exists^*\forall^*$ & \comp{PSPACE-complete} 
(Theorem~\ref{thm:nfhef.nonemptiness})\\
& $\exists^*\forall^{*\star}$ & \comp{EXPSPACE-complete} 
(Theorem~\ref{thm:nfhef.star.nonemptiness})\\
\hline
Bounded Nonemptiness & NFH & PSPACE-complete (Theorem~\ref{thm:bounded.pspace.complete})\\
\hline

\multirow{3}{*}{Universality}& $\exists\forall\forall$ & Undecidable 
(Theorem~\ref{thm:nfh.universality})\\ 
& $\exists^*~/~\forall^*$ & \comp{PSPACE-complete} 
(Theorem~\ref{thm:nfh.universality})\\ 
& $\forall^*\exists^*$ & \comp{EXPSPACE} 
(Theorem~\ref{thm:nfh.universality})\\

\hline
\multirow{2}{*}{Finite membership}& NFH & \comp{PSPACE} 
(Theorem~\ref{thm:nfh.membership.finite})\\ 
& $O(\log(k))$ ~$\forall$ & \comp{NP-complete} 
(Theorem~\ref{thm:nfh.membership.finite})\\ 
\hline
Regular membership & \multicolumn{2}{c|}{Decidable 
(Theorem~\ref{thrm:membershipFULL})} \\
\hline
\multirow{3}{*}{Containment} & NFH & Undecidable (Theorem~\ref{thm:nfh.containment})\\ & $\exists^*\subseteq \forall^*~/~\forall^*\subseteq \exists^*$ & \comp{PSPACE-complete} (Theorem~\ref{thrm:containment})\\
& $\exists^*\forall^*\subseteq \forall^*\exists^*$ & \comp{EXPSPACE}
(Theorem~\ref{thrm:containment})\\
\hline

\end{tabular} 
\caption{Summary of results on properties of hyperregular languages.}
\vspace{-10mm}
\label{tab:results}
\end{center}
\end{table}

%% file: preliminaries.tex
\vspace{-3mm}
\section{Preliminaries}
\label{sec:prelim}

An {\em alphabet} is a nonempty finite set $\Sigma$ of {\em letters}. 
A {\em word} over $\Sigma$ is a finite  
sequence of letters from 
$\Sigma$. The {\em empty word} is denoted by $\epsilon$, and the set of all words is denoted by $\Sigma^*$. A {\em language} is a subset of $\Sigma^*$.
\stam{
The {\em concatenation} of two words $w_1 = \sigma_1\cdots \sigma_n,w_2 = \sigma'_1\cdots \sigma'_m$, is
$w_1\cdot w_2 = \sigma_1\cdots \sigma_n \sigma'_1\cdots \sigma'_m$.
The concatenation of two languages $L_1,L_2$ is the language $L_1\cdot L_2 = \{w_1\cdot w_2 \mid w_1\in L_1,w_2\in L_2\}$. 
For a language $L$ and $n\in \mathbb{N}$, the language $L^n$ is defined inductively as follows. $L^0 = \{\epsilon\}$, and for $n>0$, $L^n = L^{n-1}\cdot L$.
The {\em Kleene star} of $L$ is defined as $L^* = \bigcup_{n\in\mathbb{N}} L^n$.
}
We assume that the reader is familiar with the syntax and semantics of regular expressions (RE). We use the standard notations $\{\cdot, |, *\}$ for concatenation, union, and Kleene star, respectively, and denote the language of an RE $r$ by $\lang{r}$. 
A language $L$ is {\em regular} if there exists an RE $r$ such that $\lang{r} = L$.

\stam{
\begin{definition}
\label{def:regexp}
A {\em regular expression} over $\Sigma$ is defined inductively as follows. 
\begin{itemize}
    \item $\epsilon, \emptyset$, and $\sigma$ where $\sigma\in \Sigma$ are regular expressions.
    \item Let $r_1,r_2$ be regular expressions. Then $(r_1|r_2)$, $(r_1\cdot r_2)$, and $(r_1^*)$ are regular expressions.
\end{itemize}
\end{definition}

The semantics of regular expressions assigns every regular expression $r$ a language $\lang{r}$, as follows.
\begin{itemize}
    \item $\lang{\epsilon} = \{\epsilon\}$, $\lang{\emptyset} = \emptyset$.
    \item
    $\lang{\sigma} = \{\sigma\}$. 
    $\lang{(r_1|r_2)}= \lang{r_1}\cup \lang{r_2}$, and
    $\lang{(r_1\cdot r_2)} = \lang{r_1}\cdot \lang{r_2}$, and $\lang{(r_1^*)} = \lang{r_1}^*$.
\end{itemize}
}

\begin{definition}
\label{def:nfa}
A {\em nondeterministic finite-word automaton} (NFA) is a tuple \linebreak $A 
= \tuple{\Sigma,Q,Q_0,\delta,F}$, where $\Sigma$ is an alphabet, $Q$ is a 
nonempty finite set of {\em states}, $Q_0\subseteq Q$ is a set of {\em initial 
states}, $F\subseteq Q$ is a set of {\em accepting states}, and 
$\delta\subseteq Q\times\Sigma\times Q$ is a {\em transition relation}. 
\end{definition}

Given a word $w=\sigma_1\sigma_2\cdots \sigma_n$ over $\Sigma$, a 
{\em run of $A$ on $w$} is a sequence of states $(q_0,q_1,\ldots q_n)$, such 
that $q_0\in Q_0$, and for every $0 < i \leq n$, it holds that 
$(q_{i-1},\sigma_i, q_i)\in \delta$.
The run is {\em accepting} if $q_n\in F$. 
We say that $A$ {\em accepts} $w$ if there exists an accepting run of $A$ on $w$. 
The {\em language} of $A$, denoted $\lang{A}$, is the set of all words that $A$ accepts. 
%
%
It is well-known that a language $L$ is regular iff there exists an NFA $A$ such that $\lang{A} = L$. 

%% file: HRE.tex
\section{Hyperregular Expressions}\label{sec:hre}


\begin{definition}
\label{def:hword}
A {\em hyperword over $\Sigma$} is a set of words over $\Sigma$ and a 
{\em hyperlanguage} over $\Sigma$ is a set of hyperwords over $\Sigma$.
\end{definition}


Before formally defining hyperregular expressions, we explain the idea behind them. 
A {\em hyperregular expression} (HRE) over $\Sigma$ uses a set of {\em word variables} 
$X  =\{x_1,x_2, \ldots, x_k\}$.
When expressing a hyperword $S$, these variables are assigned words from $S$. 
An HRE $r$ is composed of a {\em quantification condition} $\alpha$ over $X$,
and an underlying RE $\hat r$, which represents word assignments to $X$. An HRE $r$ defines a hyperlanguage $\hlang{r}$.
The condition $\alpha$ defines the assignments that should be in $\lang{\hat r}$. For example, $\alpha = \exists x_1.\forall x_2$ requires that 
there exists a word $w_1\in S$ (assigned to $x_1$), such that for every word 
$w_2\in S$ (assigned to $x_2$), the word that represents the assignment $x_1 \mapsto w_1, x_2 \mapsto w_2$, is in $\lang{\hat r}$. The hyperword $S$ is in $\hlang{r}$ iff $S$ meets these conditions. 

We represent an assignment $v:X\rightarrow S$ as a {\em word assignment} $\wass{w}{v}$, which is a word over the alphabet $(\Sigma\cup\{\#\})^X$ (that is, assignments from $X$ to $\Sigma\cup\{\#\}$), where the $i$'th 
letter of $\wass{w}{v}$ represents the $k$ $i$'th letters of the words 
$v(x_1),\ldots ,v(x_k)$ (in case that the words are not of equal length, we ``pad''  the end of the shorter words with $\#$ symbols).
We represent these $k$ $i$'th letters as an assignment 
denoted $\{\sigma_{1{x_1}},\sigma_{2{x_2}},\ldots, \sigma_{k{x_k}}\}$, where $x_j$ is assigned $\sigma_j$.
For example, the assignment $v(x_1)=aa$ and $v(x_2)=abb$ is represented by the word assignment $\wass{w}{v} =  \{a_{x_1},a_{x_2}\}\{a_{x_1},b_{x_2}\}\{\#_{x_1},b_{x_2}\}$. 

\begin{definition} \label{def:hyperregexp}
A {\em hyperregular expression} is a tuple $r=\tuple{X,\Sigma,\alpha,\hat r}$, where 
$\alpha = \quant_1 x_1\  \cdots \quant_k x_k$, where $\quant_i\in \{\exists,\forall\}$ for 
every $i\in[1,k]$, and where $\hat r$ is an RE over ${\hat \Sigma} = (\Sigma\cup\{\#\})^X$. 
\end{definition}

Let $S$ be a hyperword and let $v: X\rightarrow S$ be an assignment of the word 
variables of $r$ to words in $S$. We denote by  $v[x\mapsto w]$ the assignment obtained from $v$ by assigning the word $w\in S$ to $x\in X$. We represent $v$ by $\wass{w}{v}$.  
We now define the membership condition of a hyperword $S$ in the hyperlanguage of $r$. We first define a relation $\vdash$ for $S$, $\hat r$, a quantification condition $\alpha$, and an assignment $v:X\rightarrow S$, as follows.
\begin{itemize}
    \item For $\alpha = \epsilon$, define $S\vdash _v (\alpha,\hat r)$ if $\wass{w}{v}\in\lang{\hat r}$. 

\item For $\alpha = \exists x. \alpha'$, define $S\vdash_v (\alpha, \hat r)$ if  there exists $w\in S$ s.t. $S \vdash_{v[x\mapsto w]} (\alpha',\hat r)$.

\item For $\alpha = \forall x. \alpha'$, define $S\vdash_v (\alpha,\hat r)$ if $S \vdash_{v[x \mapsto w]}  
(\alpha',\hat r)$ for every $w\in S$ .\footnote{In case that $\alpha$ begins with 
$\forall$, membership holds vacuously with an empty hyperword. We 
restrict the discussion to nonempty hyperwords.}
\end{itemize}
Since all variables are under the scope of $\alpha$, membership is independent of $v$, and so if $S\vdash (\alpha,\hat r)$, we denote $S \in \hlang{r}$. The hyperlanguage of $r$ is $\hlang{r} = \{S \mid S\in\hlang{r}\}$.

\begin{definition}\label{def:hyperreglang}
We call a hyperlanguage $\hl$ a {\em regular hyperlanguage} if there exists an HRE $r$ such that $\hlang{r} = \hl$.
\end{definition}

\stam{
We consider several fragments of HRE, which limit the structure of the quantification condition $\alpha$.
$\hrefall$ is the fragment in which $\alpha$ contains only $\forall$ quantifiers, 
and similarly, in $\hree$, $\alpha$ contains only $\exists$ quantifiers. In 
the fragment $\hreef$, $\alpha$ is of the form $\exists x_1 \cdots \exists x_i \forall 
x_{i+1}\cdots \forall x_k$.
}

\input{HRE_application}

%% file: HRE_application.tex
\subsection*{Application of HRE in Information-flow Security}
\label{sec:security}


{\em Noninterference}~\cite{gm82} requires high-secret commands to be removable without 
affecting observations of users holding low clearances:
$$
\varphi_{\mathsf{ni}} = \forall x.\exists y\{l_x, l\lambda_y\}^*,
$$
where $l$ denotes a low state and $l\lambda$ denotes a low state such that all high commands are replaced by a dummy value $\lambda$.

{\em Observational determinism}~\cite{zm03} requires that if two executions of 
a system start with low-security-equivalent events, they should remain low equivalent:
$$
\varphi_{\mathsf{od}} = \forall x.\forall y.\Big( \{l_x, l_y\}^+ \mid \{\bar{l}_x, 
\bar{l}_y\}\{\$_x, \$_y\}^* \mid \{l_x, \bar{l}_y\}\{\$_x, \$_y\}^* \mid \{\bar{l}_x, l_y\}\{\$_x, \$_y\}^*\Big)
$$
where $l$ denotes a low event, $\bar{l} \in \Sigma \setminus \{l\}$, and $\$ \in  \Sigma$. 
We note that similar policies such as {\em Boudol and Castellani’s 
noninterference}~\cite{bd02} can be formulated in the same 
fashion.
\footnote{This policy states that every two executions that start from 
bisimilar states (in terms of memory low-observability), should remain 
bisimilarly low-observable.}

{\em Generalized noninterference} (GNI)~\cite{m88} allows nondeterminism in 
the low-observable behavior, but requires that low-security outputs may 
not be altered by the injection of high-security inputs:
$$
\varphi_{\mathsf{gni}} = \forall x.\forall y.\exists z. \bigg(\{h_x, l_y, hl_z\}
\mid \{\bar{h}_x, l_y, \bar{h}l_z\} \mid \{h_x, \bar{l}_y, h\bar{l}_z\} \mid \{\bar{h}_x, \bar{l}_y, 
\bar{h}\bar{l}_z\} \bigg)^*
$$
where $h$ denotes the high-security input, $l$ denotes the low-security output, 
$\bar{l} \in \Sigma \setminus\{l\}$, and $\bar{h} \in \Sigma \setminus \{h\}$.

{\em Declassification}~\cite{ss00} relaxes noninterference by allowing leaking 
information when necessary. Some programs must reveal secret information to 
fulfill functional requirements. For example, a password checker must reveal 
whether the entered password is correct or not:
$$
\varphi_{\mathsf{dc}} = \forall x.\forall y. \{li_x,li_y\}\{pw_x, pw_y\}\{lo_x, lo_y\}^+
$$
where $li$ denotes low-input state, $pw$ denotes that the password is correct, 
and $lo$ denotes low-output states. We note that for brevity, $\varphi_{\mathsf{dc}}$ does not include behaviors where the first two events are not low or,
in the second event, the password is not valid. 

{\em Termination-sensitive noninterference} requires that for two executions 
that
 start from low-observable states, information leaks are not permitted by the termination behavior of the program (here, $l$ denotes a low state and $\$ \in \Sigma$):
\begin{align*}
    \varphi_{\mathsf{tsni}} = \forall x.\forall y.\Big( \{l_x, l_y\}\{\$_x, \$_y\}^*\{l_x, l_y\}  \mid  
\{\bar{l}_x, \bar{l}_y\}\{\$_x, \$_y\}^* \mid \\\{l_x, \bar{l}_y\}\{\$_x, \$_y\}^*  \mid  \{\bar{l}_x, l_y\}\{\$_x,\$_y\}^*\Big)
\end{align*}


%% file: NFH.tex
\section{Nondeterminsitic Finite-Word Hyperautomata}
\label{sec:nfh}

We now present a  model for regular hyperlanguages, namely {\em finite-word 
hyperautomata}. A  hyperautomaton is 
composed of a set $X$ of word variables, a quantification condition, and an underlying 
finite-word automaton that accepts representations of assignments to $X$. 

\begin{definition}
\label{def:nfh}
A {\em nondeterministic finite-word hyperautomaton} (NFH) is a tuple \linebreak
$\A = \tuple{\Sigma,X,Q,Q_0,F,\delta,\alpha}$, where $\Sigma, X$ and $\alpha$ are as in Definition~\ref{def:hyperregexp}, and where $\tuple{{\hat\Sigma},Q,Q_0,F,\delta}$ forms an underlying NFA over $\hat\Sigma = (\Sigma\cup\{\#\})^X$. 
\end{definition}
The acceptance condition for NFH, as for HRE, is defined with respect to a hyperword $S$, the NFH $\A$, the quantification condition $\alpha$, and an assignment $v:X\rightarrow S$. 
For the base case of $\alpha = \epsilon$, we define $S \vdash _v (\alpha,\A)$ if
$\hat\A$ accepts $\wass{w}{v}$. The cases where $\alpha$ is of the type $\exists x. \alpha'$ and $\forall x. \alpha'$ are defined similarly as for HRE, and if $S\vdash (\alpha,\A)$, we say that {\em $\A$ accepts $S$}. 

\begin{definition}
Let $\A$ be an NFH. The {\em hyperlanguage} of $\A$, denoted $\hlang{\A}$, is 
the set of all hyperwords that $\A$ accepts.
\end{definition}

\stam{
\begin{itemize}
    \item For $\alpha = \epsilon$, we denote $S \models _v (\alpha,\A)$ if
$\hat\A$ accepts $\wass(w,v)$. 

\item For $\alpha = \exists x_i \alpha'$, we denote $S\models_v (\alpha,\A)$ if 
there exists $w\in S$, such that $S \models_{v[x_i\rightarrow w]}  
(\alpha',\A)$.

\item For $\alpha = \forall x_i. \alpha'$, we denote $S\models_v (\alpha,\A)$ 
if for every $w\in S$, it holds that $S \models_{v[x_i\rightarrow w]}  
(\alpha',\A)$.\footnote{In case that $\alpha$ begins with 
$\forall$, satisfaction holds vacuously with an empty hyperword. We 
restrict the discussion to nonempty hyperwords.}

\end{itemize}
Since the quantification condition of $\A$ includes all of $X$, acceptance is independent of the 
assignment $v$, and we denote $S \models \A$, in which case, we say that {\em $\A$ 
accepts $S$}.
}

\begin{figure}[t]
\centering
\scalebox{.8}{
        \includegraphics[scale=0.6]{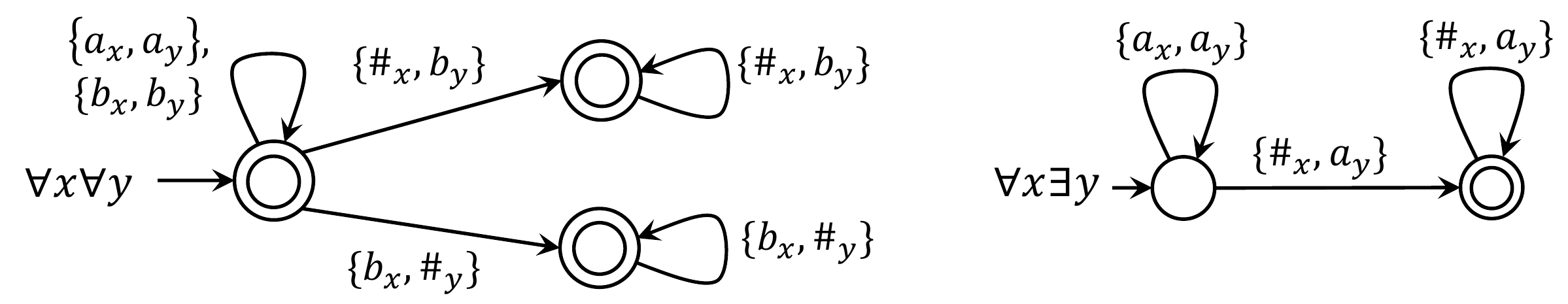}
    }
    \caption{The NFH $\A_1$ (left) and $\A_2$ (right).}
    \label{fig:nfh_examples}
\end{figure}

\begin{example}

Consider the NFH $\A_1$ in Figure~\ref{fig:nfh_examples} (left), whose alphabet 
is $\Sigma = \{a,b\}$, over two word variables $x$ and $y$. The NFH $\A_1$ contains an underlying standard NFA $\hat\A_1$. For two words $w_1,w_2$ that are assigned to $x$ and $y$, respectively, $\hat\A_1$ requires that (1) $w_1,w_2$ agree on their 
$a$ (and, consequently, on their $b$) positions, and (2) once one of the words has ended 
(denoted by $\#$), the other must only contain $b$ letters. Since the 
quantification condition of $\A_1$ is $\forall x_1. \forall x_2$, in a hyperword $S$ that is 
accepted by $\A_1$, every two words agree on their $a$ positions. As a result, all the 
words in $S$ must agree on their $a$ 
positions. The hyperlanguage of $\A_1$ is then all hyperwords in which all 
words agree on their $a$ positions. 

\end{example}

\begin{example}
The NFH $\A_2$ of Figure~\ref{fig:nfh_examples} (right) depicts the translation of the HRE of Example~\ref{example:intro_1} to an NFH.  
\end{example}


\stam{
\begin{example}
The NFH $\A_1$ and $\A_2$ in Figure~\ref{fig:nfh_examples} depict the translation of the HRE of Example~\ref{example:intro_1} and Example~\ref{example:intro_2}
\end{example}
}


\stam{
\begin{example}
Consider the NFH $\A_3$ in Figure~\ref{fig:ordered}, over the alphabet $\Sigma = 
\{a,b\}$ and two word variables $x_1$ and $x_2$. From the initial state, two 
words lead to the left component in $\hat{\A_3}$ iff in every position, if the 
word assigned to $x_2$ has an $a$, the word assigned to $x_1$ has an $a$. In 
the right component, the situation is dual -- in every position, if the word 
assigned to $x_1$ has an $a$, the word assigned to $x_2$ has an $a$. 
Since the quantification condition of $\A_3$ is $\forall x_1.\forall x_2$, in a hyperword $S$ accepted by $\A_3$, in every two words in $S$, the set of $a$ positions of one is a subset of the $a$ positions of the other. Therefore, $\hlang{\A_3}$ includes all hyperwords in which there is a full ordering on the $a$ positions.

\begin{figure}[t]
    \begin{center}
        \includegraphics[scale=0.48]{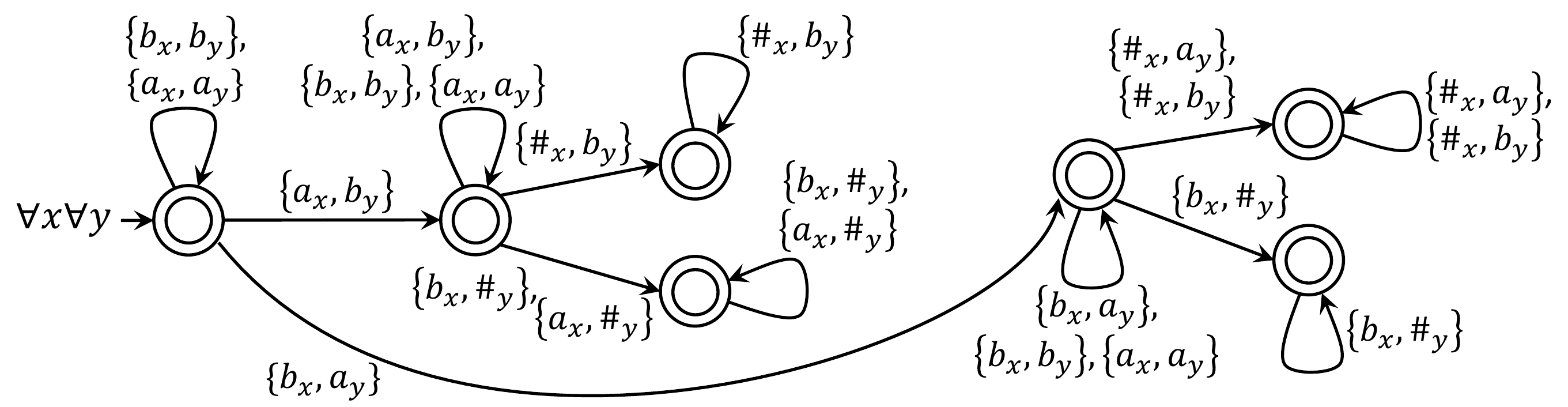}
    \end{center}
    \caption{The NFH $\A_3$.}
    \label{fig:ordered}
\end{figure}
\end{example}

}


Since regular expressions are equivalent to NFA, we can translate the underlying regular expression $\hat r$ of an HRE $r$ to an equivalent NFA, and vice versa -- translate the underlying NFA $\hat \A$ of an NFH $\A$ to a regular expression. It is then easy to see that every HRE has an equivalent NFH over the same set of variables with the same quantification condition.

We consider several fragments of NFH, which limit the structure of the quantification condition $\alpha$.
$\hrefall$ is the fragment in which $\alpha$ contains only $\forall$ quantifiers, 
and similarly, in $\hree$, $\alpha$ contains only $\exists$ quantifiers. In 
the fragment $\hreef$, $\alpha$ is of the form $\exists x_1 \cdots \exists x_i \forall x_{i+1}\cdots \forall x_k$.

\subsection{Additional Terms and Notations}

We present several terms and notations which we use throughout the paper.
Recall that we represent an assignment $v:X\rightarrow S$ as a word assignment $\wass{w}{v}$.
Conversely, a word ${\bi w}$ over $(\Sigma\cup\{\#\})^X$ represents an assignment $\assw{v}{w}:X\rightarrow \Sigma^*$, where $\assw{v}{w}(x_i)$ is formed by concatenating the letters of $\Sigma$ that are assigned to $x_i$ in the letters of ${\bi w}$.
We denote the set of all such words $\{\assw{v}{w}(x_1),\ldots,\assw{v}{w}(x_k)\}$ by $S({\bi w})$.
Since we only allow padding at the end of a word, if a padding occurs in the middle of $\bi w$, then $\bi w$ does not represent a legal assignment. Notice that this occurs iff ${\bi w}$ contains two consecutive letters ${\bi w}_i{\bi w}_{i+1}$ such that ${\bi w}_i(x) = \#$ and ${\bi w}_{i+1}(x)\neq \#$ for some $x\in X$.  
We call ${\bi w}$ {\em legal} if $\assw{v}{w}$ represents a legal assignment from $X$ to $\Sigma^*$.

Consider a function $g:A\rightarrow B$ where $A,B$ are some sets. 
The {\em range} of $g$, denoted $\range(g)$ is the set $\{g(a)| a\in A\}$.

A {\em sequence} of $g$ is a function $g':A\rightarrow B$ such that $\range(g')\subseteq \range(g)$. 
A {\em permutation} of $g$ is a function $g':A\rightarrow B$ such that $\range(g') = \range(g)$.
We extend the notions of sequences and permutations to word assignments. Let ${\bi w}$ be a word over $\hat\Sigma$. A sequence of ${\bi w}$ is a word ${\bi w'}$ such that $S(w')\subseteq S(w)$, and a permutation of ${\bi w}$ is a word ${\bi w'}$ such that $S(w')=S(w)$.

Throughout the paper, when we use a general NFH $\A$, we assume that its ingredients are as in Definition~\ref{def:nfh}.  
 \stam{
 
Consider a tuple $t = (t_1,t_2,\ldots t_k)$ of items. 
A {\em sequence} of $t$ is a tuple $(t'_1, t'_2,\ldots t'_k)$, where 
$t'_i\in\{t_1,\ldots t_k\}$ for every $1\leq i \leq k$. A {\em permutation} of $t$ is a reordering of the elements of $t$. 
We extend these notions to zipped words, to 
assignments, and to hyperwords, as follows. Let $\zeta = 
(i_1,i_2,\ldots i_k)$ be a sequence (permutation) of $(1,2,\ldots, k)$.
\begin{itemize}
    \item Let ${\bi w} = \zip(w_1,\ldots w_k)$ be a word over $k$-tuples. The word ${\bi w}_\zeta$, 
defined as $\zip(w_{i_1}, w_{i_2}, \ldots w_{i_k})$ is a sequence (permutation) of ${\bi w}$.
    \item Let $v$ be an assignment from a set of variables $\{x_1,x_2,\ldots 
x_k\}$ to a hyperword $S$. The assignment $v_\zeta$, defined as $v_\zeta(x_j) = 
v(x_{i_j})$ for every $1\leq i,j \leq k$, is a sequence (permutation) of $v$. 
\item Let $S$ be a hyperword. The tuple ${\bi w} = (w_1,\ldots w_k)$, where $w_i\in S$, is a sequence of $S$. if $\{w_1,\ldots w_k\} = S$, then $\bi w$ is a permutation of $S$. 
\end{itemize}
}

%% file: nfh_properties.tex
\section{Closure Properties of Regular Hyperlanguages}
\label{sec:nfh_closure}


We now consider closure properties of regular hyperlanguages. We show, via constructions on NFH, that regular hyperlanguages are closed under all the Boolean operations.

\begin{theorem}\label{thm:nfh.complementation}
Regular hyperlanguages are closed under complementation.
\end{theorem}

\begin{proof}
Let $\A$ be an NFH. The NFA $\hat \A$ 
can be complemented with respect to its language over $\hat\Sigma$ to an NFA 
$\overline{\hat{\A}}$. 
Then, for every assignment $v:X\rightarrow S$, it holds that $\hat \A$ accepts 
$\wass{w}{v}$ iff $\overline{\hat{\A}}$ does not accept $\wass{w}{v}$.
Let $\overline{\alpha}$ be the quantification condition obtained from $\alpha$ 
by replacing every $\exists$ with $\forall$ and vice versa. 
We can prove by induction on $\alpha$ that $\overline{\A}$, the NFH whose underlying NFA is $\overline{\hat{\A}}$, and whose quantification 
condition is 
$\overline{\alpha}$, accepts $\overline {\hlang{\A}}$.
The size of $\overline{\A}$ is exponential in $|\hat\A|$, due to the complementation construction for $\hat\A$ and complementing the set of transitions in $\delta$.
\end{proof}

\begin{theorem}\label{thm:nfh.union}
Regular hyperlanguages are closed under union.
\end{theorem}

\begin{proof}

let $\A_1 = 
\tuple{\Sigma,X,Q,Q_0,\delta_1,F_1,\alpha_1}$ and $\A_2= 
\tuple{\Sigma,Y,P,P_0,\delta_2,F_2,\alpha_2}$ be two NFH with $|X|=k$ and 
$|Y|=k'$ variables, respectively.

We construct an NFH 
$\A_{\cup} = \tuple{\Sigma,X\cup Y, Q\cup P\cup\{p_1,p_2\},  Q_0\cup P_0, \delta, 
F_1\cup F_2\cup\{p_1,p_2\}, \alpha}$, where $\alpha = \alpha_1\alpha_2$ (that is, we concatenate the two quantification conditions), and where $\delta$ is defined as follows. 
\begin{itemize}
\item For every $(q_1\xrightarrow{f} q_2)\in 
\delta_1$ we set $(q_1\xrightarrow{f\cup g} q_2)\in \delta$ for every $g\in (\Sigma\cup\{\#\})^{Y}$. 

\item
For every $(q_1\xrightarrow{f} q_2)\in \delta_2$ we set $(q_1\xrightarrow{f\cup g} q_2)\in \delta$ for every $g\in (\Sigma\cup\{\#\})^{X}$. 

\item
For every $q\in F_1$, we set 
$(q\xrightarrow{\{\#\}^X\cup g}p_1),(p_1\xrightarrow{\{\#\}^X\cup g}p_1)\in \delta$
for every $g\in (\Sigma\cup\{\#\})^{Y}$.

\item
For every $q\in F_2$, we set $(q\xrightarrow{g\cup\{\#\}^{Y}}p_2),(p_2\xrightarrow{g\cup\{\#\}^{Y}}p_2)\in \delta$ for every $g\in (\Sigma\cup\{\#\})^{X}$.

\end{itemize}

Let $S$ be a hyperword. 
For every $v:(X\cup Y)\rightarrow S$, it holds that if $\wass{w}{v|_X}\in\lang{\hat\A_1}$, then $\wass{w}{v}\in\lang{\hat\A_{\cup}}$. Indeed, according to our construction, every word assigned to the $Y$ variables is accepted in the $\A_1$ component of the construction, and so it satisfies both types of quantifiers.
A similar argument holds for $v|_Y$ and $\A_2$.

Also, according to our construction, for every $v:(X\cup Y)\rightarrow S$,  if $\wass{w}{v}\in\lang{\hat\A_{\cup}}$, then either $\wass{w}{v|_X}\in\lang{\hat\A_1}$, or $\wass{w}{v|_Y}\in\lang{\hat\A_2}$. 
As a conclusion, we have that $\hlang{\A_{\cup}} = \hlang{\A_1}\cup\hlang{\A_2}$. 

The state space of $\A_\cup$ is linear in the state spaces of $\A_1,\A_2$. 
However, the size of the alphabet of $\A_\cup$ may be exponentially larger than 
that of $\A_1$ and $\A_2$, since we augment each letter with all functions from $Y$ to $\Sigma\cup\{\#\}$ (in $\A_1$) and from $X$ to $\Sigma\cup\{\#\}$ (in $\A_2$).
\end{proof}

\begin{theorem}\label{thm:nfh.intersection}
Regular hyperlanguages are closed under intersection.
\end{theorem}

\begin{proof}
The proof follows from the closure of regular hyperlanguages under union and complementation. However, we also offer a direct translation, which avoids the need to complement. 

let $\A_1 = 
\tuple{\Sigma,X,Q,Q_0,\delta_1,F_1,\alpha_1}$ and $\A_2= 
\tuple{\Sigma,Y,P,P_0,\delta_2,F_2,\alpha_2}$ be two NFH with $|X|=k$ and 
$|Y|=k'$ variables, respectively.

We construct an NFH 
$\A_\cap = \tuple{\Sigma,X\cup Y, (Q\cup \{q\}) \times (P\cup \{p\}), (Q_0\times P_0), \delta, 
(F_1\cup\{q\}) \times (F_2\cup\{p\}), \alpha_1\alpha_2}$, where $\delta$ is defined as follows. 
\begin{itemize}
    \item For every $(q_1\xrightarrow {f} q_2)\in 
\delta_1$ and every $(p_1\xrightarrow {g} p_2)\in\delta_2$, we have 
$$\Big((q_1,p_1)\xrightarrow{f\cup g} (q_2,p_2)\Big)\in\delta$$
\item For every $q_1\in F_1, (p_1 \xrightarrow{g} p_2)\in \delta_2$ we have
$$\Big((q_1,p_1)\xrightarrow {\{\#\}^X \cup g}(q,p_2)\Big), \Big((q,p_1)\xrightarrow {\{\#\}^k\cup g}(q,p_2)\Big) \in \delta$$
%

%

\item
For every $(q_1\xrightarrow {f} q_2)\in \delta_1$ and $p_1\in F_2$, we have
$$\Big((q_1,p_1)\xrightarrow{f\cup\{\#\}^Y}(q_2,p)\Big), \Big((q_1,p)\xrightarrow{f\cup\{\#\}^Y}(q_2,p)\Big)\in \delta$$

\end{itemize}
Intuitively, the role of $q,p$ is to keep reading $\{\#\}^X$ and $\{\#\}^Y$ after 
the word read by $\hat\A_1$ or $\hat\A_2$, respectively, has ended. 

The NFH $\hat{\A_\cap}$ simultaneously reads two word assignments 
that are read along $\hat\A_1$ and 
$\hat\A_2$, and accepts iff both word assignments are accepted. The correctness 
follows from the fact that for $v:(X\cup Y)\rightarrow S$, we have that 
$\wass{w}{v}$ is accepted by $\hat\A$ iff $\wass{w}{v|_X}$ and $\wass{w}{v|_Y}$ are 
accepted by $\hat\A_1$ and $\hat\A_2$, respectively. 
This construction is polynomial in the sizes of $\A_1$ and $\A_2$.
\end{proof}

%% file: nonemptiness.tex
\section{Nonemptiness of NFH.}\label{sec:nonemptiness}

The {\em nonemptiness problem} is to decide, given an NFH $\A$, whether $\hlang{\A} = \emptyset$. 
The complexity of the nonemptiness problem affects the complexity of various other decision problems, such as universality and containment. 
In this section, we extensively study various versions of this problem for various fragments of NFH. First, we show that the problem for general NFH is undecidable. Then, we show that nonemptiness is decidable for various fragments of NFH, with varying complexities. 

We then study the {\em bounded nonemptiness} problem, in which we ask whether an NFH accepts a hyperword of bounded size.  

Finally, we study the nonemptiness problem in the presence of {\em wild-card letters}, which represent free assignments to a variable. Wild-card letters can exponentially decrease the number of transitions of an NFH. We show that for the alternation-free fragments of NFH, wild-card letters do not increase the complexity of the nonemptiness problem, while for the fragment of $\nfhef$, the smaller representation comes with an exponential blow-up in complexity.  

\subsection{General Nonemptiness Results}

We begin with the nonemptiness problem for general NFH. 

\begin{theorem}\label{thm:nfh.nonemptiness}
The nonemptiness problem for NFH is undecidable. 
\end{theorem}

\begin{proof}
In \cite{fh16}, a reduction from  the {\em 
Post correspondence problem} is used for proving the undecidability of HyperLTL satisfiability. We mimic the proof ideas of \cite{fh16} to show that the nonemptiness problem for NFH is, in general, undecidable. 
A PCP instance is a collection $C$ of dominoes of the form:
$$ \Bigg\{\Big[\frac{u_1}{v_1} \Big], 
\Big[\frac{u_2}{v_2} \Big],\dots, \Big[\frac{u_k}{v_k} \Big] \Bigg\}$$
where for all $i \in [1, k]$, we have $v_i,u_i \in 
\{a,b\}^*$.
%
The problem is to decide whether there exists a finite sequence of the dominoes 
of the form
$$\Big[\frac{u_{i_1}}{v_{i_1}} \Big]\Big[\frac{u_{i_2}}{v_{i_2}} \Big] \cdots 
\Big[\frac{u_{i_m}}{v_{i_m}} \Big]$$
where each index $i_j$ is in $[1, k]$, such that the upper and lower finite strings of the dominoes are equal, i.e.,
$$ u_{i_1}u_{i_2}\cdots{}u_{i_m} = v_{i_1}v_{i_2}\cdots{}v_{i_m}$$
For example, if the set of dominoes is
$$ C_{\mathsf{exmp}} = \Bigg\{ \Big[\frac{ab}{b}\Big], 
\Big[\frac{ba}{a}\Big],\Big[\frac{a}{aba}\Big] \Bigg\} $$
Then, a possible solution is the following sequence of dominoes from 
$C_{\mathsf{exmp}}$: 
$$\mathsf{sol} = \Big[\frac{a}{aba}\Big]\Big[\frac{ba}{a}\Big] 
\Big[\frac{ab}{b}\Big ].  $$

Given an instance $C$ of PCP, we encode a solution as a word $w_{sol}$ over the 
following alphabet:
$$\alphabet = \Big\{\frac{\sigma}{\sigma'} \mid \sigma,\sigma'\in\{a,b,{\dot 
a},{\dot b}, \$\}\Big\}.$$
Intuitively, $\dot{\sigma}$ marks the beginning of a new domino, and 
$\$$ marks the end of a sequence of the upper or lower parts of the dominoes 
sequence.


We note that $w_{sol}$ encodes a legal solution iff the following conditions are met:

\begin{enumerate}
    \item For every $ \frac{\sigma}{\sigma'}  $ that occurs in $w_{sol}$, it 
holds that $\sigma,\sigma'$ represent the same domino letter (both $a$ or both 
$b$, either dotted or undotted).
    \item The number of dotted letters in the upper part of $w_{sol}$ is equal 
to the number of dotted letters in the lower part of $w_{sol}$.
    \item $w_{sol}$ starts with two dotted letters, and the word $u_i$ between 
the $i$'th and $i+1$'th dotted letters in the upper part of $w_{sol}$, and the 
word $v_i$ between the corresponding dotted letters in the lower part of 
$w_{sol}$ are such that $ [\frac{u_i}{v_i}]  \in C$, for every $i$.
\end{enumerate}

We call a word that represents the removal of the first $k$ dominoes from 
$w_{sol}$ 
a {\em partial solution}, denoted by $w_{sol,k}$.
Note that the upper and lower parts of $w_{sol,k}$ are not necessarily of equal 
lengths (in terms of $a$ and $b$ sequences), since the upper and lower parts of 
a domino may be of different lengths, and so we use letter $\$$ to pad 
the end of the encoding in the shorter of the two parts. 

We construct an NFH $\A$, which, intuitively, expresses the following ideas: 
$(1)$ There exists an encoding $w_{sol}$ of a solution to $C$, and $(2)$ For 
every $w_{sol,k}\neq \epsilon$ in a hyperword $S$ accepted by $\A$, the word 
$w_{sol,k+1}$ is also in $S$. 

$\hlang{\A}$ is then the set of all hyperwords that contain an encoded solution 
$w_{sol}$, as well as all its suffixes obtained by removing a prefix of dominoes 
from $w_{sol}$. This ensures that $w_{sol}$ indeed encodes a legal solution. For example, a matching hyperword $S$ (for the solution 
$\mathsf{sol}$ discussed earlier) that is accepted by $\A$ is:
$$ S = \{  w_{sol} = \frac{\dot a}{\dot a}   \frac{\dot b}{b}   \frac{a}{a}   
\frac{\dot a}{\dot a}   \frac{b}{\dot b} ,
w_{sol,1} = \frac{\dot b}{\dot a}\frac{a}{\dot b}\frac{\dot a}{\$}\frac{b}{\$}, 
w_{sol,2} = \frac{\dot a}{\dot b}\frac{b}{\$}, w_{sol,3} = \epsilon
\} $$

Thus, the quantification condition of $\A$ is $\alpha = \forall x_1\exists x_2 
\exists x_3$, where $x_1$ is to be assigned a potential partial solution 
$w_{sol,k}$, and $x_2$ is to be assigned $w_{sol,k+1}$, and $x_3$ is to be 
assigned $w_{sol}$.  

During a run on a hyperword $S$ and an assignment $v:\{x_1,x_2,x_3\}\rightarrow 
S$, the NFH $\A$ checks that the upper and lower letters of $w_{sol}$ all match.
In addition, $\A$ checks that the first domino of $v(x_1)$ is indeed in $C$, and 
that $v(x_2)$ is obtained from $v(x_1)$ by removing the first tile. 
$\A$ performs the latter task by checking that the upper and lower parts of 
$v(x_2)$ are the upper and lower parts of $v(x_1)$ that have been ``shifted'' 
back appropriately. That is, if the first tile in $v(x_2)$ is the encoding of 
$[\frac{w_i}{v_i}]$, then $\A$ uses states to remember, at each point, the last 
$|w_i|$ letters of the upper part of $v(x_2)$ and the last $|v_i|$ letters of 
the lower part of $v(x_2)$, and verifies, at each point, that the next letter in 
$v(x_1)$ matches the matching letter remembered by the state.
\end{proof}

Next, we show that for the alternation-free fragments, a simple reachability test suffices to decide nonemptiness. 

\begin{theorem}\label{thm:nfhe.nfha.nonemptiness}
The nonemptiness problem for $\nfhe$ and $\nfhf$ is \comp{NL-complete}.
\end{theorem}

\begin{proof}
The lower bound for both fragments follows from the \comp{NL-hardness} of the \linebreak nonemptiness problem for NFA. 

We turn to the upper bound, and begin with $\nfhe$. Let $\A_\exists$ be an $\nfhe$. 
We claim that $\A_\exists$ is nonempty iff $\hat\A_\exists$ accepts some legal 
word ${\bi w}$. The first direction is trivial. For the second direction, let 
${\bi w}\in\lang{\hat\A_\exists}$. By 
assigning $v(x_i) = \assw{v}{w}(x_i)$ for every $x_i\in X$, we get $\wass{w}{v} = {\bi w}$, and 
according to the semantics of $\exists$, we have that $\A_\exists$ accepts $S({\bi w})$. 
To check whether $\hat\A_\exists$ accepts a legal word, we can run a 
reachability check on-the-fly, while advancing from a letter $\sigma$ to the 
next letter $\sigma'$ only if $\sigma'$ assigns $\#$ to all variables for which $\sigma$ assigns $\#$. 
While each transition $T = q\xrightarrow{f} p$ in 
$\hat\A$ is of size $k$, we can encode $T$ as 
 a set of size $k$ of encodings of 
transitions of type $q \xrightarrow {(x_i,\sigma_i)} p$ with a binary encoding of 
$p,q,\sigma_i$, as well as $i,t$, where $t$ marks the index of $T$ within the 
set of transitions of $\hat\A$. 
Therefore, the reachability test can be performed within space that is logarithmic in the size of $\A_\exists$.

Now, let $\A_\forall$ be an $\nfhf$ over $X$. We claim that $\A_\forall$ is 
nonempty iff $\A_\forall$ accepts a hyperword of size $1$. 
For the first direction, let $S\in\hlang{\A_\forall}$. Then, by the semantics of 
$\forall$, we have that for every assignment $v:X\rightarrow S$, it holds that 
$\wass{w}{v}\in\lang{\hat{\A_\forall}}$. Let $u\in S$, and let $v_u(x_i) = u$ for every 
$x_i\in X$. Then, in particular, $\wass{w}{v_u}\in\lang{\hat{\A_\forall}}$. Then for every assignment $v:X\rightarrow \{u\}$ (which consists of the 
single assignment $v_u$), it holds that $\hat{\A_\forall}$ accepts $\wass{w}{v}$, 
and therefore $\A_\forall$ accepts $\{u\}$. 
The second direction is trivial. 

To check whether $\A_\forall$ accepts a hyperword of size $1$, we restrict the reachability test on $\hat\A_\forall$ to letters over $\hat\Sigma$ that represent fixed functions. 
\end{proof}

For $\nfhef$, we show that the problem is decidable, by checking the nonemptiness of an exponentially larger equi-empty NFA.

\begin{theorem}\label{thm:nfhef.nonemptiness}
The nonemptiness problem for $\nfhef$ is  \comp{PSPACE-complete}.
\end{theorem}

\begin{proof}

Let $\A$ be an $\nfhef$ with $k$ quantifiers and $m$ $\exists$-quatifiers. We begin with a \comp{PSPACE} upper bound. 

\stam{
Let $\A$ be an $\nfhef$ with $m$ existential quantifiers, and 
let $S\in\hlang{\A}$. Then, there exist $w_1,\ldots, w_m \in S$, such that for every 
assignment 
$v:X\rightarrow S$ in which $v(x_i) = w_i$ for every $1\leq i\leq m$, we have that $\hat\A$ accepts $\wass{w}{v}$. 
In particular, $\hat\A$ accepts every assignment that agrees with $v$ on  $x_1,\ldots 
x_m$, and assigns only words from $\{w_1,\ldots, w_m\}$.
Therefore, $\hat\A$ accepts the hyperword $\{w_1,\ldots, w_m\}$. 
That is, $\A$ is nonempty iff it accepts a hyperword of size at most $m$.
We can construct an NFA $A$ based on $\hat\A$ that is nonempty iff $\hat\A$ accepts all appropriate assignments of a hyperword of size $m$. The size of $A$ is exponential in the size of $\hat\A$, and the result follows from the \comp{NL} upper bound for NFA nonemptiness. 
}

Let $S\in\hlang{\A}$. Then, according to the semantics of the quantifiers, 
there exist $w_1,\ldots w_m \in S$, such that for every assignment 
$v:X\rightarrow S$ in which $v(x_i) = w_i$ for every $1\leq i\leq m$, it holds that $\hat\A$ accepts $\wass{w}{v}$. Let $v:X\rightarrow S$ be such an 
assignment. Then, $\hat\A$ accepts 
$\wass{w}{v'}$ for every sequence $v'$ of $v$ that agrees with $v$ on its assignments to $x_1,\ldots,x_m$, and in particular, for such sequences whose range is $\{w_1,\ldots,w_m\}$. 
Therefore, by the semantics of the quantifiers, we have that $\{w_1,\ldots,w_m\}$ is 
in $\hlang{\A}$. The second direction is trivial.

We call $\wass{w}{v'}$ as described above a {\em witness to the nonemptiness of 
$\A$}.
%
We construct an NFA $A$ based on $\hat\A$ that is nonempty iff $\hat\A$ accepts 
a witness to the nonemptiness of $\A$.

Let $\Gamma$ be the set of all functions of the type $\zeta:[1,k]\rightarrow [1,m]$ such that $\zeta(i)=i$ for every $i\in[1,m]$, and such that $\range(\zeta) = [1,m]$.
For a letter assignment $f = \{\sigma_{1_{x_1}},\ldots \sigma_{k_{x_k}}\}$, we denote by $f_\zeta$ the letter assignment $\{\sigma_{\zeta(1)_{x_1}},\ldots, \sigma_{\zeta(k)_{x_k}}\}$. 

For every function $\zeta\in\Gamma$, we construct 
an NFA $A_\zeta = \tuple{ \hat\Sigma,Q,Q_0,\delta_\zeta,F}$, where for every 
$q\xrightarrow{g} q'$ in $\delta$, 
we have $q\xrightarrow{f} q'$ in 
$\delta_\zeta$, for every $f$ that occurs in $\hat\A$ for which $f_\zeta = g$.
Intuitively, for every run of $A_{\zeta}$ on a word ${\bi w}$ there exists a similar of $\hat\A$ on the sequence of ${\bi w}$ that matches $\zeta$. Therefore, $\hat\A$ accepts a witness ${\bi w}$ to the nonemptiness of $\A$ iff ${\bi w}\in\lang{A_\zeta}$ for every $\zeta\in\Gamma$. 

We define $A = \bigcap_{\zeta\in\Gamma} A_\zeta$.
Then $\hat\A$ accepts a witness to the nonemptiness of $\A$ iff $A$ is 
nonempty. 

Since $|\Gamma| = m^{k-m}$, the state space of $A$ is of size $O(n^{m^{k-m}})$, where $n=|Q|$, 
and its alphabet is of size $|\hat\Sigma|$. 
Notice that for $\A$ to be nonempty, $\delta$ must be of size at least 
$|(\Sigma\cup {\#})|^{(k-m)}$, to account for all the sequences of letters in 
the words assigned to the variables under $\forall$ quantifiers (otherwise, we 
can immediately return ``empty''). Therefore, $|\hat\A|$ is $O(n\cdot 
|\Sigma|^k)$. 
We then have that the size of $A$ is $O(|\hat \A|^k)$.  
If the number $k-m$ of $\forall$ quantifiers is fixed, then $m^{k-m}$ is 
polynomial in $k$. However, now $|\hat\A|$ may be polynomial in $n,k$, and $|\Sigma|$, 
and so in this case as well, the size of $A$ is $O(|\hat A|^k)$. 

Since the nonemptiness problem for NFA is \comp{NL-complete}, the problem for $\nfhef$ 
can be decided in space of size that is polynomial in $|{\hat\A}|$. 

For the lower bound, we show a reduction from a polynomial version of the {\em corridor tiling problem}, defined as follows. 
We are given a finite set $T$ of tiles, two relations $V \subseteq T \times T$ 
and $H \subseteq T \times T$,
an initial tile $t_0$, a final tile $t_f$, and a bound $n>0$.
We have to decide whether there is some $m>0$ and a tiling of a $n \times m$-grid such that
(1) The tile $t_0$ is in the bottom left corner and the tile $t_f$ is in the top 
right corner,
(2) A horizontal condition: every pair of horizontal neighbors is in $H$, and
(3) A vertical condition: every pair of vertical neighbors is in $V$.
When $n$ is given in unary notation, the problem is known to be 
\comp{PSPACE-complete}.
Given an instance $C$ of the tiling problem, we construct an $\nfhef$ $\A$ that is nonempty iff $C$ has a solution. 
We encode a solution to $C$ as a word $w_{sol} =w_1\cdot w_2\cdot w_m\$$ over 
$\Sigma = T\cup\{1,2,\ldots n,\$\}$, where the word $w_i$, of the form $1\cdot t_{1,i}\cdot 2 \cdot t_{2,i},\ldots n\cdot 
t_{n,i}$, describes the contents of row $i$. 

To check that $w_{sol}$ indeed encodes a solution, we need to make sure that:
\begin{enumerate}
\item $w_1$ begins with $t_0$ and $w_m$ ends with $t_f\$$.
\item $w_i$ is of the correct form.
\item Within every $w_i$, it holds that $(t_{j,i},t_{j+1,i})\in H$.
\item For $w_i,w_{i+1}$, it holds that $(t_{j,i}, t_{j,i+1})\in V$ for every $j\in[1,n]$.
\end{enumerate} 

Verifying items $1-3$ is easy via an NFA of size $O(n|H|)$.  
The main obstacle is item $4$. 

We describe an $\nfhef$ $\A = \tuple{T\cup \{0,1,2,\ldots n,\$\}, 
\{y_1,y_2,y_3,x_1,\ldots x_{\log(n)}\}$, $Q, \{q_0\},\delta, F, \alpha}$ that is 
nonempty iff there exists a word that satisfies items $1-4$.
The quantification condition $\alpha$ is $\exists y_1\exists y_2 \exists y_3\forall 
x_1 \ldots \forall x_{\log(n)}$.
The NFH $\A$ only proceeds on letters whose assignments to $y_1,y_1,y_3$ is $r,0,1$, respectively, where $r\in T\cup\{1,\ldots n,\$\}$. Notice that this means that $\A$
requires the existence of the words $0^{|w_{sol}|}$ and $1^{|w_{sol}|}$ (the $0$-word and $1$-word, henceforth).
$\A$ makes sure that the word assigned to $y_1$ matches a correct solution 
w.r.t. items $1-3$ described above.  
We proceed to describe how to handle the requirement for $V$. 
We need to make sure that for every position $j$ in a row, the tile in position $j$ in the next row matches the current one w.r.t. $V$. We can use a state $q_j$ to remember the tile in position $j$, and compare it to the tile in the next occurrence of $j$. The problem is avoiding having to check all positions 
simultaneously, which would require exponentially many states. To this end, we use $\log(n)$ copies of the $0$- and $1$-words to form a binary encoding of the position $j$ that is to be remembered. The $\log(n)$ $\forall$ conditions make sure that every position within $1-n$ is checked.  

We limit the checks to words in which $x_1,\ldots x_{\log(n)}$ are the $0$- or $1$-words, by having $\hat\A$ accept every word in which there is a letter that is not over $0,1$ that is assigned to the $x$ variables. This takes care of accepting all cases in which the word assigned to $y_1$ is also assigned to one of the $x$ variables. 

To check that $x_1,\ldots x_{\log(n)}$ are the $0$- or $1$-words, $\hat\A$ checks 
that the letter assignments to these variables remain constant throughout the run. 
In these cases, upon reading the first letter, $\hat\A$ remembers the value $j$ that is encoded by the constant assignments to $x_1,\ldots x_{\log(n)}$ in a state, and makes sure that throughout the run, the tile that occurs in the assignment to $y_1$ in position $j$ in the current row matches the tile in position $j$ in the next row. 

We construct a similar reduction for the case that the number of $\forall$ 
quantifiers is fixed: instead of encoding the position by $\log(n)$ bits, we can 
directly specify the position by a word of the form $j^*$, for every $j\in[1,n]$. 
Accordingly, we construct an $\nfhef$ over $\{x, y_1,\ldots y_{n},z\}$, with a 
quantification condition $\alpha = \exists x\exists y_1 \ldots \exists 
y_{n}\forall z$. The NFA $\hat\A$ advances only on letters whose assignments to 
$y_1,\ldots y_n$ are always \linebreak $1,2,\ldots n$, respectively, and checks only words assigned to $z$ that are some constant $1\leq j\leq n$. Notice that the fixed assignments to the $y$ variables leads to $\delta$ of polynomial size.  
In a hyperword accepted by $\A$, the word assigned to $x$ is $w_{sol}$, and the word assigned to $z$ specifies which index should be checked for conforming to $V$.
\end{proof}

\subsection{Bounded nonemptiness}
\label{sec:bounded}

The {\em bounded nonemptiness problem} is to decide, given an NFH $\A$ and $m\in \naturals$, whether $\A$ accepts a hyperword of size at most $m$.
Notice that some nonempty NFH only accept infinite hyperwords (for example, $\A_2$ of Figure~\ref{fig:nfh_examples}), and so they do not accept a hyperword of size $m$, for every $m\in\naturals$. 

We show that the bounded nonemptiness problem is decidable for all of NFH.

\begin{theorem}\label{thm:bounded.pspace}
The bounded nonemptiness problem for NFH is in \comp{PSPACE}.
\end{theorem}

\begin{proof}
Let $\A$ be an NFH with a quantification condition $\alpha$ with $k$ quantifiers, and let $m\in\naturals$.
Intuitively, we construct an NFA $A$ 
in which a single run simultaneously follows all runs of $\hat\A$ on the possible assignments of a potential hyperword $S$ of size $m$ to the variables of $\A$. 
Then, $A$ accepts a set of such legal assignments (represented as a single word) iff $\A$ accepts a hyperword of size at most $m$.

The {\em assignment tree} for $\alpha$ and $m$ is defined as follows. The tree $T$ has $k+1$ levels, where the root is at level $0$.  For $0<i\leq k$, if $\quant_i = \forall$, then every node in level $i-1$ has $m$ children. If $\quant_i = \exists$, then every node in level $i-1$ has a single child. 
Every node $v$ in $T$ is associated with an encoding in $[1,m]^*$ that matches the path from the root to $v$. For example, if $v$ is in level $2$, and $\alpha$ begins with $\exists\forall$, and $v$ is the second child, then the position of $v$ is encoded by $1\cdot 2$. The leaves of $T$ are then all encoded by elements of $[1,m]^k$. 

A {\em labeling} $c$ of $T$ labels every node (except for the root) by some value in $[1,m]$. For $0<i\leq k$, if $\quant_i = \forall$, then the $m$ children of every node in level $i-1$ are labeled $1$ to $m$.
If $\quant_i = \exists$, then the child of every node in level $i-1$ is labeled by some value in $[1,m]$. 

Consider a hyperword $S=\{w_1,w_2,\ldots w_m\}$. 
Every path $p$ along $c(T)$ matches an assignment of the words in $S$ to the variables in $X$: the variable $x_i$ is assigned $w_j$, where $j$ is the labeling of the node in level $i$ in $p$. Then, $c(T)$ matches a possible set of assignments of the words of $S$ to the variables in $X$. Given $p$, we denote this assignment by $f_p$.
According to the semantics of NFH, we have that $\A$ accepts a hyperword of size $m$ iff there a labeling $c(T)$ such that for every path $p$ of $c(T)$, the underlying NFA $\hat\A$ accepts the word assignment for $f_p$.

We construct $A$ such that a single run of $A$ simultaneously follows every assignment $f_p$ in a labeling $c(T)$, letter by letter.

Let $C$ be the set of all labelings of $T$, and let $L$ be the set of all indices of leaves of $T$. We define the NFA $A$ as follows.   
The alphabet of $A$ is $(\Sigma\cup\{\#\})^m$.
The set of states of $A$ is $Q^{L}\times C$. The set of initial states is $Q_0^L\times C$, and the set of accepting states is $F^L\times C$.

The transition relation of $A$ is as follows. We add a transition labeled $(\sigma_1,\sigma_2,\ldots \sigma_m)$ from
$(((q_1,l_1),\ldots (q_{|L|}, l_{|L|})),c)$ to $(((q'_1,l_1),\ldots (q'_{|L|}, l_{|L|})),c')$ if $c=c'$, and for every $1\leq r\leq |L|$, there is a transition in $\delta$ labeled by $\{\sigma'_{1_{x_1}},\sigma'_{2_{x_2}},\ldots \sigma'_{k_{x_k}}\}$ from  $q_r$ to $q'_r$, where $\sigma'_{i_{x_i}} = \sigma_j$, where $j$ is the labeling of the node in level $i$ in the path to $l_r$ in $c$. 

For example, consider the labeled assignment tree $c(T)$ of Figure~\ref{fig:tree.assign} for the quantification condition $\forall x_1 \exists x_2$, and $m=3$. 
Then $T$ has three leaves, labeled $1\cdot 1$, $2\cdot 1$, and $3
\cdot 1$. The labeling $c(T)$ assigns the nodes of $T$ values in $[1,3]$ as described in Figure~\ref{fig:tree.assign}.  
The three transitions in $\A$ from $q_1,q_2,q_3$ are then translated to the transition from $s = (((q_1,1\cdot1),(q_2,2\cdot 1),(q_3,3\cdot1)), c(T))$ labeled $(a,b,c)$, which means that the transition associates label $1$ with $a$, label $2$ with $b$, and label $3$ with $c$, matching the transitions from $q_1$,$q_2$, and $q_3$, when they are associated with the leaves as in $s$. 

\begin{figure}[t]
\centering
\scalebox{.8}{
        \includegraphics[scale=0.6]{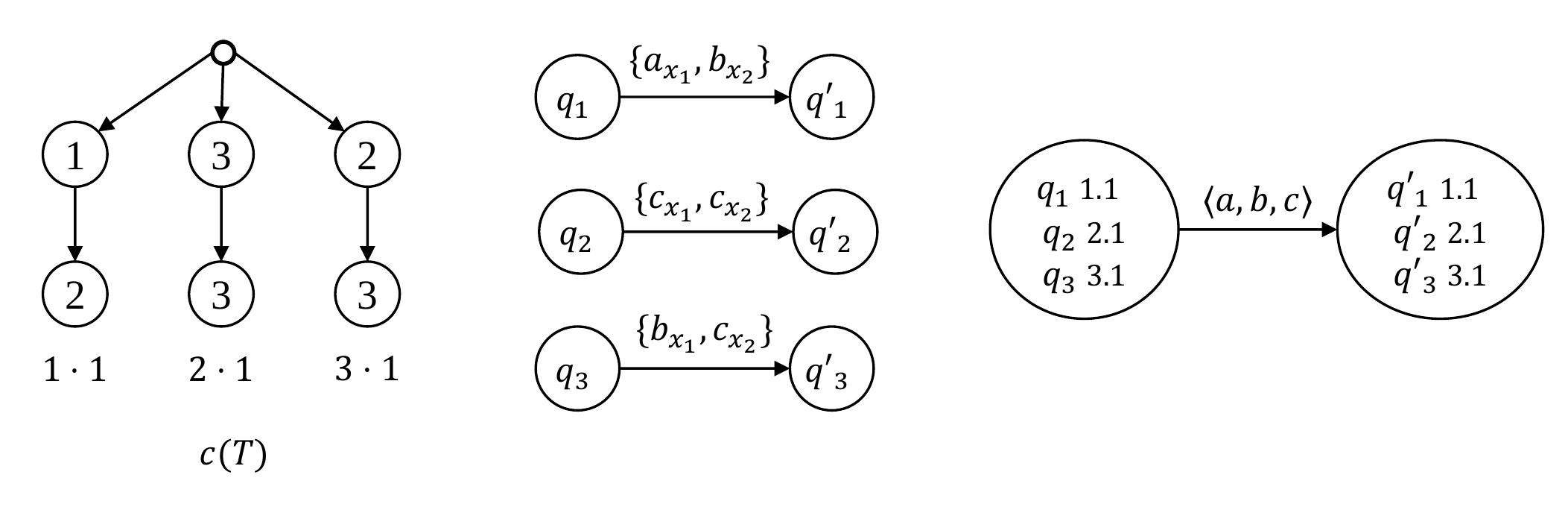}
    }
    \caption{The labeled assignment tree $c(T)$ (left), transitions in $\A$ (middle), and their depiction in $A$ (right).}
    \label{fig:tree.assign}
\end{figure}

The size of $T$ (and hence, the size of $L$) is $O(m^{k'})$, where $k'$ is the number of $\forall$ quantifiers in $\alpha$. Accordingly, the size of $C$ is $O(m^{m^{k'}})$.
Therefore, the state space of $A$ is of size $O(n^{m^{k'}}\cdot m^{m^{k'}})$, where $n$ is the number of states in $\A$.  

According to our construction, we have that $\A$ accepts a hyperword of size $m$ iff $A$ is nonempty, when considering only paths that are legal assignments, that is, once a value $i$ is assigned the letter $\#$, it continues to be assigned $\#$. 
Checking $A$ for such nonemptiness can be done on-the-fly in space that is logarithmic in the size of $A$. 
Notice, as mentioned in the proof of theorem~\ref{thm:nfhef.nonemptiness}, that for $m>1$, the size of the transition relation of $\A$ must be exponential in the size of $k'$, to account for the different assignments to the $\forall$-quantifiers (otherwise, $\A$ is empty and we can return ``false'').
Therefore, the size of each state of $A$ is polynomial in the size of $\A$, and a \comp{PSPACE} upper bound follows.  
\end{proof}

A \comp{PSPACE} lower bound for the bounded nonemptiness problem for NFH directly follows from the nonemptiness problem for $\nfhef$, since, as we prove in Theorem~\ref{thm:nfhef.nonemptiness}, an $\nfhef$ $\A$ with $k'$ $\exists$-quantifiers is nonempty iff it accepts a hyperword of size $k'$. However, we prove \comp{PSPACE-hardness} for a $\forall x \exists y$ quantification condition, showing that this problem is \comp{PSPACE-hard} even for a fixed number of $\forall$ and $\exists$ quantifiers. 

\begin{theorem}\label{thm:bounded.pspace.hard}
The bounded nonemptiness problem for NFH with $\alpha = \forall x \exists y$ is \comp{PSPACE-hard}.
\end{theorem}
\begin{proof}
We reduce from the problem of deciding the nonemptiness of the intersection of $k$ given NFA, which is known to be \comp{PSPACE-hard}.

Let $A_1,A_2,\ldots A_k$ be NFA, where $A_i = \tuple{\Sigma, Q_i, Q_0^i, \delta_i, F_i}$. We construct an NFH $\A = \tuple{\Sigma',\{x,y\},Q,Q_0,F,\delta,\forall x\exists y}$ that accepts a hyperword whose size is at most $k$ iff there exists a word $w$ such that $w\in \lang{A_i}$ for every $i\in[1,k]$. 

The set of states $Q$ of $\A$ is $\bigcup_i Q_i\times Q_{(i+1)\textsf{mod} k}$, and $\Sigma' = \bigcup_i Q_i\times\Sigma\times Q_i$. The set of accepting states is $\bigcup_i F_i\times F_{(i+1)\textsf{mod} k}$, and the set of initial states $Q_0$ is $\bigcup_i Q_0^i\times Q_0^{(i+1)\textsf{mod} k}$. 
The transitions are as follows. 
For every $i\in[1,k]$, every $\sigma\in\Sigma$, and every two transitions $(q,\sigma,q')\in\delta_i, (p,\sigma,p')\in\delta_{(i+1)\textsf{mod} k}$, we set $((q,p),\{(q,\sigma,q')_x,(p,\sigma,p')_y\},(q',p'))\in\delta$. 
Notice that the size of $\A$ is polynomial in the sizes of $A_1,\ldots A_k$. 
Every word assignment $\wass{w}{}$ that is read along $\hat\A$ describes the parallel run of $A_i$ and $A_{(i+1)\textsf{mod} k}$ on the same word $w$. The word assignment $\wass{w}{}$ is accepted by $\hat\A$ iff $w$ is accepted by both $A_i$ and $A_{(i+1)\textsf{mod} k}$. 

If there exists a word $w$ that is accepted by all NFA, then the hyperword $S$ that describes all the matching accepting runs on $w$ by the different NFA is accepted by $\A$. Indeed, for the accepting run on $w$ by $A_i$ there is a matching accepting run on $w$ by $A_{(i+1)\textsf{mod} k}$.

Conversely, if there exists a hyperword of size (at most) $k$ that is accepted by $\A$, then it contains descriptions of runs of $A_1,\ldots A_k$ on words. By the way we have defined $\A$, if there exists $r\in S$ that describes the accepting run of $A_i$ on a word $w$, then there must exist  $r'\in S$ that describes the accepting run of $A_{(i+1)\textsf{mod} k}$ on $w$. As a result, and combined with the size of $S$, we have that $S$ must contain an accepting run of every NFA in the set, and these runs must all be on the same word $w$. Therefore, the intersection of $A_1,\ldots A_k$ is nonempty.  
\end{proof}

As a conclusion from Theorems~\ref{thm:bounded.pspace} and \ref{thm:bounded.pspace.hard}, we have the following.

\begin{theorem}\label{thm:bounded.pspace.complete}
The bounded nonemptiness problem for NFH is \comp{PSPACE-complete}. 
\end{theorem}

\input{nfh_wildcard}

\subsection{A semi-algorithm for deciding the nonemptiness for $\forall\exists$}
\label{sec:semialgo}

The nonemptiness problem for NFH is undecidable already for the fragment of $\forall\exists$, as shown in Theorem~\ref{thm:nfh.nonemptiness}. However, this fragment is of practical use in expressing finite-word properties, as shown in Section~\ref{sec:security}. 
We now describe a semi-algorithm for testing the nonemptiness of an NFH with a quantification condition of the type $\forall\exists$.
Intuitively, this procedure aims at finding the largest hyperword that is accepted by the NFH. 

The procedure first considers the set $L_0$ of all the words that can be assigned to $x_1$, and checks whether this set subsumes the matching assignments for the $\exists$ quantifier. If so, then $L_0$ is a suitable hyperword. Otherwise, $L_0$ is pruned to the largest potential hyperword by omitting from $L_0$ all words that are not assigned to the variable under $\exists$, and the procedure continues to the next round. In case that the procedure does not find an accepted hyperword, or conversely if the procedure does not reach an empty set, it does not halt.   

We describe our procedure with more detail. 
Let $\A = \tuple{\Sigma,\{x,y\},Q,Q_0,F,\delta,\forall x\exists y}$ be an NFH. 
Let $L^0_\forall = \{u|\exists v: \wass{w}{x\mapsto u, y\mapsto v}\in \lang{\hat\A} \}$, and let $L^0_\exists =\{ v|\exists u: \wass{w}{x\mapsto u, y\mapsto v}\in \lang{\hat\A}$\}. 
We denote the NFA obtained from $\hat\A$ by restricting the transitions to assignments to $x$ by $\hat\A_x$, and similarly define $\hat\A_y$.
It is easy to see that $A^0_\forall = \hat\A_x$ is an NFA for $L^0_\forall$, and $A^0_\exists =\hat\A_y$ is an NFA for $L^0_\exists$. 

If $L^0_\exists\subseteq L^0_\forall$, then by the semantics of NFH, we have that $L^0_\forall$ is accepted by $\A$.
If $L^0_\exists\cap L^0_\forall = \emptyset$, then by the semantics of NFH, we have that $\A$ is empty.
Otherwise, there exists a word in $L^0_\exists$ that is not in $L^0_\forall$, and vice versa.

We define $L^1_\exists = L^0_\exists\cap L^0_\forall$. Notice that $L^1_\exists$ is regular, and an NFA $A^1_\exists$ for $L^1_\exists$ can be calculated by the intersection construction for $A^0_\forall$ and $A^0_\exists$. Now $L^1_\exists\subseteq L^0_\forall$. However, it may be the case that there exists a word $u\in L^0_\forall$ for which there exists no matching $v\in L^1_\exists$. 
Therefore, we restrict $L^0_\forall $ to a set $L^1_\forall = \{u|\exists v\in L^1_\exists: \wass{w}{x\mapsto u,y\mapsto v}\in\lang{\hat\A}\}$.  
We calculate an NFA $A^1_\forall$ for $L^1$, as follows.
Let 
$A^0_\exists = \tuple{P, \Sigma,p_0, \delta_0, F_0}$. 
We define
$\hat A^1 = \tuple{Q\times P, (\Sigma\cup\{\#\})^{\{x,y\}}, (q_0,p_0),\delta_1, F_1\times F_2}$, where $\delta_1 = \{((q,p),\{\sigma'_x,\sigma_y\}, (q',p') | \sigma,\sigma'\in\Sigma,  (q,\{\sigma_x,\sigma'_y\},q')\in\delta,(p,\sigma',p')\in\delta_0\}$. That is, $\hat\A^1$ is roughly the intersection construction of $\hat\A$ and $A^0_\exists$, when considering only the letter assignments to $y$. We denote this construction by $\cap_y$.
Finally, we set $A^1_\forall = \hat\A^1_x$.

Now, if $\lang{A^1_\exists}\subseteq \lang{A^1_\forall}$, then $\lang{A^1_\forall}$ is accepted by $\A$, and if $\lang{A^1_\exists}\cap \lang{A^1_\forall} = \emptyset$, then $\lang{\A} = \emptyset$. Otherwise, we repeat the process above with respect to $\hat\A^1,A^1_\forall, A^1_\exists$.

Algorithm~\ref{algorithm:ae} describes the procedure. 

  \begin{algorithm}[ht]\label{algorithm:ae}
            \SetAlgoLined
            \DontPrintSemicolon
            \KwInput{$\A$.}
            \KwOutput{$\hlang{\A}\neq\emptyset$?}
              
            $A_\forall = \hat\A_x, A_\exists = \hat\A_y$\;
            \While{true}
            {
               	\If{$\lang{A_\exists}\subseteq \lang{A_\forall}$}
               	{
               	    \textbf{return} $\true$\;
               	} \ElseIf{$\lang{A_\exists}\cap \lang{A_\forall} = \emptyset$}
               	{
               	    \textbf{return} $\false$\;
               	}
               	
               	$A_\exists = A_\exists \cap A_\forall$ \;
               	$\hat\A = \hat\A\cap_y \A_\exists$\;
               	$A_\forall = \hat\A_x $
              }
        	\textbf{endwhile}\;
            \caption{Nonemptiness test for $\forall\exists$}
        \end{algorithm}

%% file: nfh_wildcard.tex
\subsection{NFH with Wild Card Letters}
\label{sec:wildcard}

When constructing an HRE or an NFH, every letter must include an assignment to all variables. However, an HRE may only need to describe the assignment to a subset of the variables at each step. For example, the HRE $$ \exists x \exists y \{a_x\}\{b_y\} $$ describes hyperwords in which there exist two words, where the first word starts with $a$, and the second word has $b$ in its second position. Since the first letter and the second letter of the second and first words, respectively, do not matter, there is no need to express them. 
Therefore, we can define a more general and useful notion of HRE in which the letters are {\em partial} functions from $X$ to $\Sigma$. 

To translate the notion of partial functions to NFH, we simply add a wild-card letter $\star$ which can stand for every letter assignment to the variables. For example, the letter $\{a_x,\star_y\}$ stands for all the assignments to $x,y$ in which $x$ is assigned $a$. 

The size of the alphabet $\hat\Sigma$ of an underlying NFA must be exponential in the size of the number of $\forall$-quantifiers, to account for all the assignments of letters to all the variables under $\forall$-quantifiers. Otherwise, the language of the NFH is empty. Using wild-card letters, such transitions can be replaced by a single transition in which every variable under $\forall$ is assigned $\star$. 
Thus, using wild-card letters can lead to exponentially smaller NFH. 

We define NFH with wild cards accordingly. An NFH with wild card letters (NFH$^\star$) is a tuple
$\A = \tuple{ \Sigma,X,Q,Q_0,F,\delta,\alpha}$ whose underlying NFA $\hat\A$ is over the alphabet $\hat\Sigma = (\Sigma\cup\{\#,\star\})^X$.
The semantics of NFH$^\star$ is similar to that of NFH. The only difference is that now, $\wass{w}{v}$ contains all possible word assignments in which the letters in $\Sigma$ may also be replaced with $\star$ in the assignments to the variables. 

Obviously, every NFH$^\star$ can be translated to an NFH with an exponential blow-up in the number of transitions. The constructions for intersection, union, and complementation can all be adjusted to handle the wild cards. Due to the exponential decrease in size, the complexity of the various decision procedures for NFH$^\star$ may, in the worst case, increase exponentially. Since the nonemptiness problem is at the core of most decision procedures, we study its complexity for the various fragments of NFH$^\star$.

We begin with $\nfhe^\star$ and $\nfhf^\star$, and show that for these fragments, adding wild-card letters does not change complexity of the nonemptiness problem.  

According to the proof of Theorem~\ref{thm:nfhe.nfha.nonemptiness}, a simple reachability test on the underlying NFA suffices to determine nonemptiness for these fragments. We notice that this holds also in the presence of wild-card letters. Indeed, an $\nfhf^\star$ is nonempty iff it accepts a hyperword of size $1$. The proof of Theorem~\ref{thm:nfhe.nfha.nonemptiness} locates such a word by following an accepting path in the underlying NFH in which all variables are equally assigned at every step. It is easy to see that such a path suffices also when some of the variables are assigned wild-card letters. Similarly, an accepting path in an $\nfhe$ induces a finite accepted hyperword, and the same holds also when traversing transitions with wild-card letters.
Therefore, we have the following. 

\begin{theorem}\label{thm:nfhe.nfha.star.nonemptiness}
The nonemptiness problem for $\nfhe^\star$ and $\nfhf^\star$ is \comp{NL-complete}.
\end{theorem}

We turn to study the fragment of $\nfhef^\star$. Recall that in the proof for the lower bound of Theorem~\ref{thm:nfhef.nonemptiness}, we argue that the size of the transition relation of a nonempty $\nfhef$ must be exponential in its number of $\forall$-quantifiers, which affects the space complexity analysis of the size of the NFA that we construct. 
For an $\nfhef^\star$ $\A$, this argument no longer holds. While we can construct a similar NFA and check its nonemptiness, its size may now be exponential in that of $\A$, conforming to an \comp{EXPSPACE} upper bound. We prove a matching lower bound, and conclude that in contrast to the alternation-free fragments, adding wild-card letters hardens the nonemptiness problem for $\nfhef$. 

\begin{theorem}\label{thm:nfhef.star.nonemptiness}
The nonemptiness problem for $\nfhef^\star$ is \comp{EXPSPACE-complete}. 
\end{theorem}

\begin{proof}
Let $\A$ be an $\nfhef$. 
Consider the NFA $A$ constructed in the proof of Theorem~\ref{thm:nfhef.nonemptiness}. A similar NFA can be constructed to decide the nonemptiness of $\A$. The only difference is the need to consider the intersection of letters which carry wild-card letters. These can be easily computed: the intersection letter of $\{\sigma_{1_{x_1}},\sigma_{2_{x_2}},\ldots \sigma_{k_{x_k}}\}$ and $\{\sigma'_{1_{x_1}},\sigma'_{2_{x_2}},\ldots \sigma'_{k_{x_k}}\}$ is $\{\gamma_{1_{x_1}},\gamma_{2_{x_2}},\ldots \gamma_{k_{x_k}}\}$, where $\gamma_i = \sigma_i$ if $\sigma'_i =\star$, and  $\gamma_i = \sigma'_i$ if $\sigma_i=\star$, and otherwise it must hold that $\gamma_i = \sigma_i = \sigma'_i$.

The size of $A$ is, as in the proof of Theorem~\ref{thm:nfhef.nonemptiness}, $O(n^{m^{k-m}})$, where $n$ is the number of states in $\A$, and $m$ is the number of $\exists$-quantifiers in $\alpha$. 
Since the nonemptiness problem for NFA is \comp{NL-complete}, an \comp{EXPSPACE} upper bound follows. 

We turn to the lower bound. As in the proof of Theorem~\ref{thm:nfhef.nonemptiness}, we reduce from the corridor tiling problem: we are given an input $C$ which consists of a finite set $T$ of tiles, two relations $V \subseteq T \times T$ 
and $H \subseteq T \times T$,
an initial tile $t_0$, a final tile $t_f$, and a bound $n>0$.
In the exponential version of this problem, we need to decide whether there exists a legal tiling of a $2^n\times m$ for some $m>0$ (in contrast to the polynomial version which we use for $\nfhef$). 
This problem is known to be \comp{EXPSPACE-complete}. 

We use a similar idea as for $\nfhef$, and encode the legal solution as a word, while using the $0-$ and $1-$words under $\forall$ as memory. However, the exponential length of each row in the tiling poses two main obstacles. First, we can no longer use a state to remember the index in the row that we need to check in order to verify the vertical condition. Second, we can no longer use numbered letters to mark the index in every row, and using binary encoding requires verifying that the encoding is correctly ordered. 
We describe how we overcome these two obstacles by using wild-card letters. 

We encode a solution $w_{sol} = \$ w_1\cdot w_2\cdot w_m\$$ over 
$\Sigma = T\cup\{0,1,\$,\&\}$, where the word $w_i$, of the form $b_0\cdot t_{0,i}\cdot b_1 \cdot t_{2,i},\ldots b_{2^n-1}\cdot 
t_{2^n-1,i}$, describes the contents of row $i$, where $b_j$ is the $n$-bit binary encoding of index $j$.
Additionally, we use the $0$-word which only consists of $0$ letters, and similarly we use the $1$-word. Here, we precede the sequence of bits with $\&$. 

We construct an $\nfhef^\star$ $\A$ with a quantification condition \linebreak $\alpha = \exists s \exists x_0 \exists x_1 \forall u \forall y_1 \ldots \forall y_n \forall z_1 \ldots \forall z_n$ that is nonempty iff $C$ has a solution. 
Intuitively, as in the proof of Theorem~\ref{thm:nfhef.nonemptiness}, the assignment to $s$ must be $w_{sol}$, and the assignment to $x_0$ and $x_1$ must be the $0$- and the $1$-words, respectively. The assignment to $u$ must be equal to the assignment of either $s$, $x_0$, or $x_1$. 
Notice that since $u$ is under $\forall$, then if $\A$ is nonempty then the only hyperword it can accept is $\{w_{sol}, 0, 1\}$. Therefore, the rest of the variables must always be assigned one of these three words in order for $\A$ to accept. 

When the assignments to $y_1\ldots y_n$ are the $0$- and $1$- words, their binary values are used for encoding a single index $j$ that verifies that every two consecutive tiles in position $j$ satisfy $V$, as we describe below. $\hat\A$ accepts all runs in which one of the $y$ variables is assigned $w_{sol}$. To this end, the transition relation $\delta$ of $\hat\A$ uses transitions from the initial state labeled by letters in which one of $y_1,\ldots y_n$ is assigned $\$$ and the rest are assigned $\star$, leading to accepting runs for these cases. 

To match the encoding of the $y$ variables with the correct index $j$ in $w_{sol}$, the transition relation $\delta$ of $\A$ describes the $n$ bits of $j$ in cycles of length $n+1$, where in each cycle, the $i$'th bit of $j$ is specified in the $i$'th step, and the rest of the values are represented as $\star$.
In each cycle, the $i$'th bit is compared with the $i$'th bit in $w_{sol}$. In cycles in which all $n$ index bits in $w_{sol}$ match those of $y_1\ldots y_n$, the tile in the letter that follows the encoding (the $n+1$'th letter in the cycle) is matched with the previous tile, remembered by a state, to verify that they satisfy $V$. 

For example, for $n=3$, the encoding $101$ would be as follows. 
$$
\begin{pmatrix}
y_1 & = & \& & 1 & \star & \star & \star & 1 & \star & \star\cdots \\
y_2 & = & \& & \star & 0 & \star & \star & \star & 0 & \star \cdots \\
y_3 & = & \& & \star & \star & 1 & \star & \star & \star & 1 \cdots 
\end{pmatrix}
$$
Notice that (considering only $y$ variables), only $2n+2$ letters are needed to describe this encoding: two for every value of the $i$'th bit, one of all wild-cards, and one for all $\&$. Specifying all bits in a single letter would require exponentially many letters.

We now describe how to verify that the index encoding along $w_{sol}$ is correct. 
We use the $z$ variables in a similar way to the $y$ variables, to encode the successor position of the one encoded in the $y$ variables. 
To check that they are indeed successors, it suffices to check, within the first cycle, that all bits up to some $1\leq i < n$ are equal, that $z_i = 1$ and $y_i = 0$, and that $y_{i+1}\ldots y_n = 1$ and $z_{i+1}\ldots z_n= 0$ (the only exception is for $2^n+1$ and $0$, in which we only need to check that all $y$ bits are $1$ and all $z$ bits are $0$).
Runs of $\hat\A$ in which the encoding in the $z$ variables is not the successor of the encoding of the $y$ variables, or in which one of the $z$ variables is assigned $w_{sol}$, are accepting. Otherwise, whenever the encoding of the position in $w_{sol}$ is equal to that of the $y$ variables (we check this bit by bit), we check that the encoding of the position in the next cycle is equal to that of the $z$ variables.

For example, for checking the successor of $101$, the assignments to the $y$ and $z$ variables would be as follows. 
$$
\begin{pmatrix}
y_1 & = & \& & 1 & \star & \star & \star & 1 & \star & \star\cdots \\
y_2 & = & \& & \star & 0 & \star & \star & \star & 0 & \star \cdots \\
y_3 & = & \& & \star & \star & 1 & \star & \star & \star & 1 \cdots \\
z_1 & = & \& & 1 & \star & \star & \star & 1 & \star & \star\cdots \\
z_2 & = & \& & \star & 1 & \star & \star & \star & 1 & \star \cdots \\
z_3 & = & \& & \star & \star & 0 & \star & \star & \star & 0 \cdots 
\end{pmatrix}
$$

Since the $y$ and $z$ variables are under $\forall$, all positions along $w_{sol}$ are checked over all runs of $\hat\A$ on the different assignments to the $y$ and $z$ variables. It is left to check that the first position in $w_{sol}$ is $0^n$, and the last position is $1^n$, which can be done via states. 

Checking the horizontal condition itself can be done by comparing every two consecutive tiles in the same row. These tiles are $n$ letters apart, and so this can be done via the states and does not require using the variables as memory. The rest of the checks, i.e, the identity of the first and last tiles, and the correct form of $w_{sol}$, can also be easily checked by the states. 

In every letter of $\hat\A$ (other than the first in the run, in which all $y$ and $z$ variables are assigned $\&$), there are at most six non-wild card letters: the assignments to $s,x_0,x_1$ and $u$, and $y_i$ and $z_i$ for some $1\leq i\leq n$, and additionally the letters in which one of the $y$ or $z$ variables is assigned with a word that starts with $\$$. Therefore, the alphabet of $\A$ is polynomial in the input. The number of states needed for the various checks is also polynomial, and therefore the size of $\A$ is polynomial in $|C|$. 
\end{proof}

%% file: decision_procedures.tex
\section{Additional decision procedures}
\label{sec:decproc}

The {\em universality problem} is to decide whether a given NFH $\A$ accepts every hyperword over $\Sigma$. Notice that $\A$ is universal iff $\overline{\A}$ is empty. Since complementing an NFH involves an exponential blow-up, we conclude the following from the results in Section~\ref{sec:nonemptiness}, combined with the PSPACE lower bound for the universality of NFA.

\begin{theorem}\label{thm:nfh.universality}
	The universality problem for 
\begin{enumerate}
    \item NFH is undecidable,
    \item $\nfhe$ and $\nfhf$ is \comp{PSPACE-complete}, and
    \item $\nfhfe$ is in \comp{EXPSPACE}.
\end{enumerate}
\end{theorem}



We turn to study the membership problem for NFH: given an NFH $\A$ and a 
hyperword $S$, is $S\in\hlang{\A}$? 
When $S$ is finite, so is the set of assignments from $X$ to $S$, 
and so the problem is decidable. We call this case the {\em finite membership 
problem}. 

\begin{theorem}\label{thm:nfh.membership.finite}
\begin{itemize}
\item The finite membership problem for NFH is in \comp{PSPACE}. 
\item
The finite membership problem for a hyperword of size $k$ and an NFH with $O(\log(k))$ $\forall$ quantifiers is \comp{NP-complete}. 
\end{itemize}
\end{theorem}

\begin{proof}
Let $S$ be a finite hyperword, and let $\A$ be an NFH with $k$ variables.
We can decide the membership of $S$ in $\hlang{\A}$ by iterating over all relevant assignments from $X$ to $S$, and for every such assignment $v$, checking on-the-fly whether $\wass{w}{v}$ is accepted by $\hat\A$. 
This algorithm uses space of size that is polynomial in $k$ and logarithmic in $|\A|$. 

In the case that the number of $\forall$ quantifiers is $O(\log k)$, an \comp{NP} upper bound is met by iterating over all assignments to the variables under $\forall$, while guessing assignments to the variables under $\exists$. For every such assignment $v$, checking whether $\wass{w}{v}\in\lang{\hat\A}$ can be done on-the-fly. 

We show \comp{NP-hardness} for this case by a reduction from the Hamiltonian cycle problem. 
Given a graph $G = \tuple{V,E}$ where $V = \{v_1,v_2,\ldots, v_n\}$ and 
$|E|=m$, we construct an $\nfhe$ $\A$ over $\{0,1\}$ with $n$ states, $n$ variables, $\delta$ of size $m$, and a hyperword $S$ of size $n$, as follows. $S = \{w_1,\ldots, w_n\}$, where $w_i$ is the word over $\{0,1\}$ in which all letters are $0$ except for the $i$'th. 
The structure of $\hat\A$ is identical to that of $G$, and we set $Q_0 = F = \{v_1\}$. For the transition relation, for every $(v_i,v_j)\in E$, we have $(v_i, \f_i,v_j)\in \delta$, where $\f_i$ assigns $0$ to all variables except for $x_i$. 
Intuitively, the $i$'th letter in an accepting run of $\hat\A$ marks traversing 
$v_i$. Assigning $w_j$ to $x_i$ means that the $j$'th step of the run 
traverses $v_i$. Since the words in $w$ make sure that every $v\in V$ is 
traversed exactly once, and that the run on them is of length $n$, we have that $\A$ accepts $S$ iff there exists some ordering of the words in $S$ that matches a Hamiltonian cycle in $G$.

\noindent{\it remark}
To account for all the assignments to the $\forall$ variables, $\delta$ -- and therefore, $\hat\A$ -- must be of size at least $2^{k'}$ (otherwise, we can return ``no'').
We then have that if $k = O(k')$, then space of size $k$ is logarithmic in $|\hat\A|$, and so the problem in this case can be solved within logarithmic space.
A matching NL lower bound follows from the membership problem for NFA. 
\end{proof}

When $S$ is infinite, it may still be finitely represented, allowing for algorithmic membership testing. 
We now address the problem of deciding whether a regular language $\cal L$ 
(given as an NFA) is accepted by an NFH. We call this {\em the regular 
membership problem for NFH}. We show that this problem is decidable for the 
entire class of NFH.

\begin{theorem}
\label{thrm:membershipFULL}
The regular membership problem for NFH is decidable.
\end{theorem}

\begin{proof}
Let $A = \tuple{\Sigma, P, P_0, \rho,F}$ be an NFA, and let $\A = \tuple{\Sigma, 
\{x_1,\ldots, x_k\}, Q, Q_0,$ $\delta, {\cal F},\alpha}$ be an NFH. 

First, we construct an NFA $A'=\tuple{\Sigma\cup\{\#\}, P', P'_0, \rho', F'}$
by extending the alphabet of $A$ to $\Sigma\cup\{\#\}$, adding a new and 
accepting state $p_f$ to $P$ with a self-loop labeled by $\#$, and transitions 
labeled by $\#$ from every $q\in F$ to $p_f$. 
The language of $A'$ is then $\lang{A}\cdot \#^*$.  
We describe a recursive procedure (iterating over $\alpha$) for deciding 
whether $\lang{A}\in\hlang{\A}$.

For the case that $k=1$, if $\alpha = \exists x_1$, then 
$\lang{A}\in\hlang{\A}$ iff $\lang{A}\cap \lang{\hat{\A}} \neq \emptyset$.
Otherwise, if $\alpha = \forall x_1$, then $\lang{A}\in\hlang{\A}$ iff 
$\lang{A}\notin \hlang{\overline{\A}}$, where $\overline{\A}$ is the NFH for $\overline{\hlang{\A}}$. 
The quantification condition for $\overline{\A}$
is $\exists x_1$, conforming to the base case.

For $k>1$, we construct a sequence of NFA $A_k,A_{k-1} \ldots, A_1$ as follows.
Initially,  $A_{k} = \hat\A$.
Let $A_i = \tuple{\Sigma_i, Q_i, 
Q^0_i, \delta_i, {\cal F}_i}$.
If $\quant_i = \exists$ , then 
we construct $A_{i-1}$ as follows. 
The set of states of $A_{i-1}$ is $Q_i\times P$, and the set of initial 
states is $Q_i^0\times P_0$. The set of accepting states is ${\cal F}_i\times F$.
For every 
$(q\xrightarrow{f}q')\in\delta_i$ and every 
$(p\xrightarrow{f(x_i)}p')\in \rho$, we have 
$((q,p)\xrightarrow{f\setminus{\{\sigma_{i_{x_i}}}\}}(q',p'))\in\delta_{i-1}$. We denote this construction by $A\cap_{x_i} A_i$.
Then, $A_{i-1}$ accepts a word assignment $\wass{w}{v}$ iff there 
exists a word $u\in \lang{A}$, such that $A_{i}$ accepts 
$\wass{w}{v\cup\{x_i\mapsto u\}}$.


If $\quant_i = \forall$, then  we set $A_{i-1} = \overline{A\cap_{x_i} \overline{A_i}}$
Notice that $A_{i-1}$ accepts a word assignment $\wass{w}{v}$ iff for every $u\in \lang{A}$, it holds that $A_{i}$ accepts 
$\wass{w}{v\cup\{x_i\mapsto u\}}$. 

For $i\in [1,k]$, let $\A_{i}$ be the NFH whose quantification condition is \linebreak $\alpha_{i} = \quant_1 x_1\cdots \quant_{i} x_{i}$, and whose underlying NFA is $A_i$. Then, according to the construction of $A_{i-1}$, we have that $\lang{A}\in\hlang{\A_i}$ iff $\lang{A}\in\hlang{\A_{i-1}}$.

The NFH $\A_1$ has a single variable, and we can now apply the base case.

Every $\forall$ quantifier requires complementation, which is exponential in $|Q|$.
Therefore, in the worst case, the complexity of this algorithm is $O(2^{2^{...^{|Q||A|}}})$, where the tower is of height $k$. If the number of $\forall$ quantifiers is fixed, then the complexity is $O(|Q||A|^k)$. 
\end{proof}


The {\em containment problem} is to decide, given NFH $\A_1$ and $\A_2$, whether $\hlang{\A_1}\subseteq \hlang{\A_2}$. Since we can reduce the nonemptiness problem to the containment problem, we have the following as a result of Theorem~\ref{thm:nfh.nonemptiness}.
\begin{theorem}\label{thm:nfh.containment}
The containment problem for NFH is undecidable.
\end{theorem}

However, the containment problem is decidable for various fragments of NFH. 

\begin{theorem}
\label{thrm:containment}
    The containment problem of $\nfhe\subseteq \nfhf$ and $\nfhf\subseteq\nfhe$ is \comp{PSPACE-complete}. 
    The containment problem of $\nfhef\subseteq\nfhfe$ is in \comp{EXPSPACE}
\end{theorem}

\begin{proof}
A lower bound for all cases follows from the \comp{PSPACE-hardness} of the containment problem for NFA. 
For the upper bound, for two NFH $\A_1$ and $\A_2$, we have that 
$\hlang{\A_1}\subseteq\hlang{\A_2}$ iff 
$\hlang{\A_1}\cap\overline{\hlang{\A_2}}  =  \emptyset$. 
We can compute an NFH $\A = 
\A_1\cap\overline{\A_2}$ (Theorems~\ref{thm:nfh.complementation},~\ref{thm:nfh.intersection}), and check its nonemptiness. Complementing $\A_2$ is exponential in its number of states, and the intersection construction is polynomial. 

If $\A_1\in\nfhe$ and $\A_2\in\nfhf$ or vice versa, then $\A$ is an $\nfhe$ or 
$\nfhf$, respectively, whose nonemptiness can be decided in space that is 
logarithmic in $|\A|$.   

The quantification condition of an NFH for  the intersection 
may be any 
interleaving of the quantification conditions of the two intersected NFH. (Theorem~\ref{thm:nfh.intersection}).
Therefore, for the rest of the fragments, we can construct the intersection such that $\A$ is an $\nfhef$. 
The exponential blow-up in complementing $\A_2$, along with The \comp{PSPACE} upper bound of Theorem~\ref{thm:nfhef.nonemptiness} gives an \comp{EXPSPACE} upper bound for the rest of the cases. 
\end{proof}

%% file: related.tex
\section{Related Work}
\label{sec:related}

It is well-known that classic specification languages like regular expressions and LTL cannot
express hyperproperties. The study of specific hyperproperties, such as noninterference, dates
back to the seminal work by Goguen and Meseguer \cite{gm82} in the
1980s.
The first systematic study of hyperproperties is due to Clarkson and
Schneider~\cite{cs10}.
Subsequently, temporal logics HyperLTL and HyperCTL* were introduced~\cite{cfkmrs14} to 
give formal syntax and semantics to hyperproperties. HyperLTL was recently extended to 
A-HLTL~\cite{bcbfs21} to capture {\em asynchronous} hyperproperties, where some execution traces 
can stutter while others advance.

There has been much recent progress in automatically
{verifying}~\cite{frs15,fmsz17,fht18,cfst19,hsb21}
and {monitoring}~\cite{ab16,fhst19,bsb17,bss18,fhst18,sssb19,hst19} 
HyperLTL specifications. HyperLTL is also supported by a growing set of 
tools, including the model checkers HyperQube~\cite{hsb21}, MCHyper~\cite{frs15,cfst19}, the  
satisfiability checkers EAHyper~\cite{fhs17} and MGHyper~\cite{fhh18}, and the 
runtime monitoring tool RVHyper~\cite{fhst18}.

Related to the nonemptiness problem in this paper is the \emph{satisfiability} problem for 
HyperLTL, which was shown to be decidable for the $\exists^*\forall^*$ fragment, and undecidable for any fragment 
that includes a $\forall\exists$ quantifier alternation~\cite{fh16}. The hierarchy of hyperlogics beyond 
HyperLTL has been studied in~\cite{cfhh19}. Furthermore, our other results are 
aligned with the complexity of HyperLTL model checking for tree-shaped and 
general Kripke structures~\cite{bf18}, which encode finite traces. In particular, our membership results 
are in line with the results on the complexity of verification in~\cite{bf18}. This shows that the 
complexity results in~\cite{bf18} mainly stem from the nature of quantification 
over finite words and depend on neither the full power of the temporal operators 
nor the infinite nature of HyperLTL semantics. 

The \emph{synthesis} problem has shown to be undecidable in general, and 
decidable for the $\exists^*$ and $\exists^*\forall$ fragments. While the 
synthesis problem becomes, in general, undecidable as soon as there are two 
universal quantifiers, there is a special class of universal specifications, 
called the linear $\forall^*$-fragment, which is still decidable~\cite{fhlst18}. 
The linear $\forall^*$-fragment corresponds to the decidable \emph{distributed 
	synthesis} problems. 
The \emph{bounded synthesis} problem considers only systems up to a given bound on the number of 
states.
Bounded synthesis from hyperproperties is studied 
in~\cite{fhlst18}, and has been successfully applied to 
small examples such as the dining cryptographers~\cite{c85}. Program repair and 
controller synthesis for HyperLTL have been studied in~\cite{bf19,bf20}.
Our results on bounded nonemptiness complement the 
known results, as it resembles the complexity of bounded synthesis.

%% file: discussion.tex
\section{Discussion and Future Work}
\label{sec:concl}

We have introduced and studied {\em hyperlanguages} and a framework for their modeling, 
focusing on the basic class of regular hyperlanguages, modeled by HRE and NFH. We have shown that regular hyperlanguages are closed under set operations and are capable of 
expressing important hyperproperties for information-flow security policies over 
finite traces. We have also investigated fundamental decision procedures for various fragments of NFH, conscentrating mostly on the important decision problem of nonemptiness. Some gaps, such as the precise lower bound for the universality and containment problems for $\nfhef$, are left open. 


Since our framework does not limit the type of underlying model, it can be lifted to handle hyperwords consisting of infinite words, with an underlying model designed for such languages, such as {\em nondeterministic B{\"u}chi automata}, which model $\omega$-regular languages. Just as B{\"u}chi automata can express LTL, such a model can express the entire logic 
of HyperLTL~\cite{cfkmrs14}.

\stam{
In an NBA, an infinite word is accepted iff it traverses an accepting state infinitely often. NBAs carry similar properties to NFA, with similar constructions. In particular, they are closed under the Boolean operations, their nonemptiness and membership problems are \comp{NL-complete}, and their containment problem is \comp{PSPACE-complete}. The proofs and results of Sections~\ref{sec:nfh_dp} and~\ref{sec:nfh_closure} can then be carried over also for the case of $\omega$-regular hyperlanguages and hyperautomata with an underlying NBA.

{\em HyperLTL}~\cite{cfkmrs14} is a logic for expressing hyperproperties that is based on the syntax of LTL.
While an LTL formula describes the behavior of a single trace, a hyperLTL formula describes the behavior of a set of traces, by using quantified {\em trace variables} in an LTL formula that uses these variables to describe the (mutual) behavior of the traces that are assigned to them. A hyperLTL formula is of the form $\g = \alpha\f$, where $\alpha$ includes the quantified trace variables, and $\f$ is an LTL formula. 
The Vardi-Wolper construction~\cite{VW94} translates an LTL formula $\f$ to an NBA whose language comprises all words that satisfy $\f$. 
Hyperautomata with an underlying NBA can therefore express the entire logic of hyperLTL, simply by applying the Vardi-Wolper construction for $\f$ to produce an underlying NBA, whose language exactly expresses the behavior of the traces assigned to the trace variables, along with the quantification condition $\alpha$. 
Since the translation of an LTL formula to an NBA is exponential in the size of the formula, we have that hyperLTL can be translated to hyperautomata with an underlying NBA with an exponential blow-up. 
This blow-up, along with 
the complexity results for the various decision problems, are in line with the results on satisfiability and model-checking of HyperLTL~\cite{fh16} and its 
various fragments. Thus, hyperautomata with underlying NBA are optimal, complexity-wise, for handling HyperLTL.
}

\stam{
We note that our framework can be trivially lifted to infinite words as well. 
and obtain B{\"u}chi hyperautomata that can express the entire logic 
of HyperLTL~\cite{cfkmrs14}. This can be achieved using the standard 
Vardi-Wolper construction for LTL~\cite{VW94} as basis and augmenting the 
resulting automaton with quantifiers of the input HyperLTL formula. Complexity 
results for the various decision procedures for NFH, combined with the 
complexity results shown in~\cite{fh16} would carry.
}

As future work, we plan on studying non-regular hyperlanguages 
(e.g., context-free), and object hyperlanguages (e.g., trees). Another direction is designing learning algorithms for hyperlanguages, by exploiting known canonical forms for the underlying models, and basing on existing learning algorithms for them. The main challenge would be handling learning sets and  a mechanism for learning word variables and quantifiers.

%% file: main.bbl
\begin{thebibliography}{10}
\providecommand{\url}[1]{\texttt{#1}}
\providecommand{\urlprefix}{URL }
\providecommand{\doi}[1]{https://doi.org/#1}

\bibitem{abbd20-atva}
{\'{A}}brah{\'{a}}m, E., Bartocci, E., Bonakdarpour, B., Dobe, O.:
  Probabilistic hyperproperties with nondeterminism. In: Proceedings of the
  18th Symposium on Automated Technology for Verification and Analysis (ATVA).
  pp. 518--534 (2020)

\bibitem{ab18}
{\'{A}}brah{\'{a}}m, E., Bonakdarpour, B.: {H}yper{PCTL}: {A} temporal logic
  for probabilistic hyperproperties. In: QEST. pp. 20--35 (2018)

\bibitem{ab16}
Agrawal, S., Bonakdarpour, B.: Runtime verification of $k$-safety
  hyperproperties in {H}yper{LTL}. In: Proceedings of the {IEEE} 29th Computer
  Security Foundations (CSF). pp. 239--252 (2016)

\bibitem{as85}
Alpern, B., Schneider, F.: Defining liveness. Information Processing Letters
  pp. 181--185 (1985)

\bibitem{bf18}
B.~Bonakdarpour, B., Finkbeiner, B.: The complexity of monitoring
  hyperproperties. In: CSF. pp. 162--174 (2018)

\bibitem{bcbfs21}
Baumeister, J., Coenen, N., Bonakdarpour, B., S{\'{a}}nchez, B.F.C.: A temporal
  logic for asynchronous hyperproperties. In: Proceedings of the 33rd
  International Conference on Computer-Aided Verification (CAV) (2021), to
  appear

\bibitem{bf19}
Bonakdarpour, B., Finkbeiner, B.: Program repair for hyperproperties. In:
  Proceedings of the 17th Symposium on Automated Technology for Verification
  and Analysis (ATVA). pp. 423--441 (2019)

\bibitem{bf20}
Bonakdarpour, B., Finkbeiner, B.: Controller synthesis for hyperproperties. In:
  Proceedings of the {IEEE} 32th Computer Security Foundations (CSF). pp.
  366--379 (2020)

\bibitem{bss18}
Bonakdarpour, B., S{\'{a}}nchez, C., Schneider, G.: Monitoring hyperproperties
  by combining static analysis and runtime verification. In: ISoLA. pp. 8--27
  (2018)

\bibitem{bs21}
Bonakdarpour, B., Sheinvald, S.: Finite-word hyperlanguages. In: Language and
  Automata Theory and Applications - 15th International Conference, {LATA}
  2021, Milan, Italy, March 1-5, 2021, Proceedings. Lecture Notes in Computer
  Science, vol. 12638, pp. 173--186. Springer (2021)

\bibitem{bsb17}
Brett, N., Siddique, U., Bonakdarpour, B.: Rewriting-based runtime verification
  for alternation-free {H}yper{LTL}. In: Proceedings of the 23rd International
  Conference on Tools and Algorithms for the Construction and Analysis of
  Systems (TACAS). pp. 77--93 (2017)

\bibitem{c85}
Chaum, D.: Security without identification: Transaction systems to make big
  brother obsolete. Communications of the {ACM}  \textbf{28}(10),  1030--1044
  (1985)

\bibitem{cfkmrs14}
Clarkson, M., Finkbeiner, B., Koleini, M., Micinski, K., Rabe, M.,
  S{\'{a}}nchez, C.: Temporal logics for hyperproperties. In: POST. pp.
  265--284 (2014)

\bibitem{cs10}
Clarkson, M., Schneider, F.: Hyperproperties. Journal of Computer Security pp.
  1157--1210 (2010)

\bibitem{cfst19}
Coenen, N., Finkbeiner, B., C.~S{\'{a}}nchez, C., Tentrup, L.: Verifying
  hyperliveness. In: CAV. pp. 121--139 (2019)

\bibitem{cfhh19}
Coenen, N., Finkbeiner, B., Hahn, C., Hofmann, J.: The hierarchy of
  hyperlogics. In: Proceedings 34th Annual {ACM/IEEE} Symposium on Logic in
  Computer Science (LICS). pp. 1--13 (2019)

\bibitem{eh86}
Emerson, E.A., Halpern, J.: ``sometimes" and ``not never" revisited: on
  branching versus linear time temporal logic. Journal of the {ACM} pp.
  151--178 (1986)

\bibitem{fht19}
Finkbeiner, B., Haas, L., Torfah, H.: Canonical representations of $k$-safety
  hyperproperties. In: CSF 2019. pp. 17--31 (2019)

\bibitem{fh16}
Finkbeiner, B., Hahn, C.: Deciding hyperproperties. In: CONCUR. pp. 13:1--13:14
  (2016)

\bibitem{fhh18}
Finkbeiner, B., Hahn, C., Hans, T.: {MGHyper}: Checking satisfiability of
  {HyperLTL} formulas beyond the {\textbackslash}exists
  {\^{}}*{\textbackslash}forall {\^{}}* {\(\exists\)} {\({_\ast}\)}
  {\(\forall\)} {\({_\ast}\)} fragment. In: Proceedings of the 16th
  International Symposium on Automated Technology for Verification and Analysis
  (ATVA). pp. 521--527 (2018)

\bibitem{fhlst18}
Finkbeiner, B., Hahn, C., Lukert, P., Stenger, M., Tentrup, L.: Synthesizing
  reactive systems from hyperproperties. In: Proceedings of the 30th
  International Confer ence on Computer Aided Verification (CAV). pp. 289--306
  (2018)

\bibitem{fhs17}
Finkbeiner, B., Hahn, C., Stenger, M.: Eahyper: Satisfiability, implication,
  and equivalence checking of hyperproperties. In: Proceedings of the 29th
  International Conference on Computer Aided Verification (CAV). pp. 564--570
  (2017)

\bibitem{fhst18}
Finkbeiner, B., Hahn, C., Stenger, M., Tentrup, L.: {RVH}yper: {A} runtime
  verification tool for temporal hyperproperties. In: Proceedings of the 24th
  International Conference on Tools and Algorithms for the Construction and
  Analysis of Systems (TACAS). pp. 194--200 (2018)

\bibitem{fhst19}
Finkbeiner, B., Hahn, C., Stenger, M., Tentrup, L.: Monitoring hyperproperties.
  Formal Methods in System Design (FMSD)  \textbf{54}(3),  336--363 (2019)

\bibitem{fht18}
Finkbeiner, B., Hahn, C., Torfah, H.: Model checking quantitative
  hyperproperties. In: Proceedings of the 30th International Conference on
  Computer Aided Verification. pp. 144--163 (2018)

\bibitem{fmsz17}
Finkbeiner, B., M\"uller, C., Seidl, H., Zalinescu, E.: Verifying {S}ecurity
  {P}olicies in {M}ulti-agent {W}orkflows with {L}oops. In: Proceedings of the
  15th ACM Conference on Computer and Communications Security (CCS) (2017)

\bibitem{frs15}
Finkbeiner, B., Rabe, M.N., S{\'{a}}nchez, C.: Algorithms for model checking
  {H}yper{LTL} and {H}yper{CTL}*. In: Proceedings of the 27th International
  Conference on Computer Aided Verification (CAV). pp. 30--48 (2015)

\bibitem{bd02}
G.~Boudol, G., Castellani, I.: Noninterference for concurrent programs and
  thread. In: TCS 2002. pp. 109--130 (2002)

\bibitem{gm82}
Goguen, J., Meseguer, J.: Security policies and security models. In: IEEE Symp.
  on Security and Privacy. pp. 11--20 (1982)

\bibitem{hst19}
Hahn, C., Stenger, M., Tentrup, L.: Constraint-based monitoring of
  hyperproperties. In: Proceedings of the 25th International Conference on
  Tools and Algorithms for the Construction and Analysis of Systems (TACAS).
  pp. 115--131 (2019)

\bibitem{hsb21}
Hsu, T.H., S{\'{a}}nchez, C., Bonakdarpour, B.: Bounded model checking for
  hyperproperties. In: Proceedings of the 27th International Conference on
  Tools and Algorithms for Construction and Analysis of Systems (TACAS). pp.
  94--112 (2021)

\bibitem{m88}
McCullough, D.: Noninterference and the composability of security properties.
  In: Proceedings of the 1988 {IEEE} Symposium on Security and Privacy. pp.
  177--186 (1988)

\bibitem{p77}
Pnueli, A.: The temporal logic of programs. In: FOCS. pp. 46--57 (1977)

\bibitem{ss00}
Sabelfeld, A., Sands, D.: Probabilistic noninterference for multi-threaded
  programs. In: CSFW. pp. 200--214 (2000)

\bibitem{sssb19}
Stucki, S., S{\'{a}}nchez, C., Schneider, G., Bonakdarpour, B.: Graybox
  monitoring of hyperproperties. In: Proceedings of the 23rd International
  Symposium on Formal Methods (FM). pp. 406--424 (2019)

\bibitem{vw86}
Vardi, M., Wolper, P.: Automata theoretic techniques for modal logic of
  programs. Journal of Computer and System Sciences pp. 183--221 (1986)

\bibitem{VW94}
Vardi, M., Wolper, P.: Reasoning about infinite computations. Information and
  Computation pp. 1--37 (1994)

\bibitem{wzbp19}
Wang, Y., Zarei, M., Bonakdarpour, B., Pajic, M.: Statistical verification of
  hyperproperties for cyber-physical systems. {ACM} Transactions on Embedded
  Computing systems (TECS) pp. 92:1--92:23 (2019)

\bibitem{zm03}
Zdancewic, S., Myers, A.: Observational determinism for concurrent program
  security. In: CSFW. p.~29 (2003)

\end{thebibliography}
